\documentclass[prd,showpacs,preprintnumbers,amsmath,amssymb]{revtex4}

\usepackage{graphics}
\usepackage{epsfig}
\usepackage{dcolumn}
\usepackage{bm}

\begin{document}
\title{$Z^+(4430)$ as a $D_1'{D}^* $ ($D_1{D}^* $) molecular state}


\author{Xiang Liu$^{1,2}$}\email{xiangliu@pku.edu.cn}
\author{Yan-Rui Liu$^{3}$}\email{yrliu@ihep.ac.cn}
\author{Wei-Zhen Deng$^{1}$}
\author{Shi-Lin Zhu$^{1}$}\email{zhusl@phy.pku.edu.cn}

\affiliation{$^1$Department of Physics, Peking University, Beijing
100871, China \\
$^2$Centro de F\'{i}sica Te\'{o}rica, Departamento de F\'{i}sica, Universidade de Coimbra,
P-3004-516, Coimbra, Portugal\\
$^3$Institute of High Energy Physics, P.O. Box 918-4, Beijing
100049, China}

\date{\today}

\begin{abstract}

We reexamine whether $Z^+(4430)$ could be a $D_1'-{D}^*$ or
$D_1-{D}^*$ molecular state after considering both the pion and
$\sigma$ meson exchange potentials and introducing the form factor
to take into account the structure effect of the interaction
vertex. Our numerical analysis with Matlab package MATSLISE
indicates the contribution from the sigma meson exchange is small
for the $D_1'-{D}^*$ system and significant for the $D_1-{D}^*$
system. The S-wave $D_1-\bar{D}^*$ molecular state with only
$J^{P}=0^-$ and $D_1'-{D}^*$ molecular states with
$J^P=0^-,1^-,2^-$ may exist with reasonable parameters. One should
investigate whether the broad width of $D_1'$ disfavors the
possible formation of molecular states in the future. The bottom
analog $Z_B$ of $Z^+(4430)$ has a larger binding energy, which may
be searched at Tevatron and LHC. Experimental measurement of the
quantum number of $Z^+(4430)$ may help uncover its underlying
structure.

\end{abstract}


\maketitle

\section{introduction}\label{sec1}

Recently Belle collaboration observed one very exotic resonance
$Z^{+}(4430)$ in the $\pi^{+}\psi'$ invariant mass spectrum in the
exclusive $B\to K \pi^{+}\psi'$ decays \cite{Belle-4430}. Its mass
and width are
\begin{eqnarray*}
M=4433\pm4(\mathrm{stat})\pm1(\mathrm{syst})\;\;
\mathrm{MeV}\end{eqnarray*} and
\begin{eqnarray*}
\Gamma=44^{+17}_{-13}(\mathrm{stat})^{+30}_{-11}(\mathrm{syst})\;\;\mathrm{MeV}.
\end{eqnarray*}
This resonance appears as an excellent candidate of either the
multiquark state or the molecular state and has stimulated many
theoretical investigations
\cite{xiangliu,rosner,Meng,Bugg,cky,maiani,Gershtein,Qiao,qsr,LiY,
ding,braaten,LiuXH,voloshin,polosa,godfrey}. A concise review of
the theoretical status of $Z^+(4430)$ can be found in Ref.
\cite{xiangliu}. Thereafter Ding studied $Z^+(4430)$ using the
approach of effective Lagrangian and predicted a $B^*B_1$ bound
state with mass 11048.6 MeV \cite{ding}. Later Braaten and Lu
studied the line shapes of $Z^+(4430)$ \cite{braaten}.

In our previous work \cite{xiangliu}, we explored whether
$Z^+(4430)$ could be a loosely bound S-wave molecular state of
$D^*$ and $\bar{D}_1'$ (or $\bar{D}_1$) with $J^P=0^-, 1^-, 2^-$.
We considered the one-pion exchange potential only and did not
introduce the form factor in the scattering matrix elements in the
derivation of the potential. Instead of solving the Schrodinger
equation numerically, we employed some simple trial wave functions
and the variation method to study whether a shallow bound state
exists. We found that the interaction from the one pion exchange
alone is not strong enough to bind the pair of charmed mesons with
realistic coupling constants. Other dynamics is necessary if
$Z^+(4430)$ is a molecular state.

The one pion exchange potential alone does not bind neutron and
proton to form the deuteron in nuclear physics either. The strong
attractive force in the intermediate range is required in order to
bind the deuteron, which is modelled by the sigma meson exchange.
Meanwhile hadrons are not point-like objects. The cutoff should be
introduced to describe the structure effect of the interaction
vertex. With the above considerations, we have reexamined the
$D\bar{D}^*$ molecule picture for $X(3872)$ after taking into
account both the pion and sigma meson exchange. It turns out that
the sigma meson exchange potential is repulsive and numerically
important \cite{3872-liu}. One may wonder whether the similar
mechanism plays a role in the case of $Z^+(4430)$. Therefore we
will make a comprehensive study whether $Z^+(4430)$ is the
molecular state.

This work is organized as follows. We present the effective
Lagrangian relevant to calculate the $\pi$ and $\sigma$ exchange
potentials for $Z^+(4430)$ in Section \ref{sec2}. We present the
detail of the derivation of the potential in Section \ref{sec3}.
We discuss the behavior of the potential in Section \ref{sec4}. We
make numerical analysis in Section \ref{sec5}. We discuss the
bottom analog of $Z^+(4430)$ in Section \ref{sec6}. We discuss the
cutoff dependence in Section \ref{sec7}. The last section is the
conclusion.

\section{The effective Lagrangian and coupling constants}\label{sec2}
In order to derive the $\pi$ and $\sigma$ exchange potentials, we
collect the relevant effective Lagrangian in this section. The
Lagrangian for the interaction of $\pi$ and charmed mesons is
constructed in chiral symmetry and heavy quark limit
\cite{falk,casalbuoni}
\begin{eqnarray}
\mathcal{L}_\pi&=&ig {\rm Tr}[H_b
{A}\!\!\!\slash_{ba}\gamma_5\bar{H}_a ]+ig'{\rm Tr}[ S_b
{A}\!\!\!\slash_{ba}\gamma_5\bar{S}_a]
 +ig''{\rm Tr}[T_{\mu b}
A\!\!\!\slash_{ba}\gamma_5\bar{T}_a^{\mu}]+ih {\rm
Tr}[S_b{A}\!\!\!\slash_{ba}\gamma_5
\bar{H}_a]\nonumber\\&&+i\frac{h_1}{\Lambda_{\chi}}{\rm
Tr}[T_b^{\mu}(D_{\mu}{A}\!\!\!\slash)_{ba}\gamma_5\bar{H}_a]
 +i\frac{h_2}{\Lambda_{\chi}}{\rm
Tr}[T_b^{\mu}(D\!\!\!\!/A_{\mu})_{ba}\gamma_5\bar{H}_a]+h.c.\label{aa}
\end{eqnarray}
where the fields $H$, $S$ and $T$ are defined in terms of the
$(0^-, 1^-)$, $(0^+, 1^+)$, $(1^+, 2^+)$ doublets respectively
\begin{eqnarray}
H_a&=&\frac{1+\not v}{2 }[P_{a}^{*\mu}\gamma_\mu-P_a \gamma_5],\\
S_{a}&=&\frac{1+\not v}{2 }[P_{1a}^{'\mu}\gamma_{\mu}\gamma_5
-P_{0a}^{*}],\\
T_{a}^{\mu}&=&\frac{1+\not v}{2 }\Big\{P^{*\mu\nu}_{2a}
\gamma_{\nu}-\sqrt{\frac{3}{2}}P_{1a}^{\nu}\gamma_5 [g_{\nu}^{\mu}
-\frac{1}{3}\gamma_{\nu}(\gamma^{\mu}-v^{\mu})]\Big\}.
\end{eqnarray}
Here the axial vector field $A_{ab}^{\mu}$ is defined as
\begin{eqnarray*}
A_{ab}^{\mu}=\frac{1}{2}(\xi^{\dag}\partial^{\mu}\xi-\xi\partial^{\mu}\xi^{\dag})_{ab}=
\frac{i}{f_{\pi}}\partial^{\mu}\mathcal{M}_{ab}+\cdots
\end{eqnarray*}
with $\xi=\exp(i\mathcal{M}/f_{\pi})$, $f_\pi=132$ MeV and
\begin{eqnarray}
\mathcal{M}&=&\left(\begin{array}{ccc}
\frac{\pi^{0}}{\sqrt{2}}+\frac{\eta}{\sqrt{6}}&\pi^{+}&K^{+}\\
\pi^{-}&-\frac{\pi^{0}}{\sqrt{2}}+\frac{\eta}{\sqrt{6}}&
K^{0}\\
K^- &\bar{K}^{0}&-\frac{2\eta}{\sqrt{6}}
\end{array}\right).
\end{eqnarray}

In Eq. (\ref{aa}), the coupling constants were estimated in the
quark model \cite{falk},
\begin{eqnarray}
g=g_A,\quad g'=g_A/3, \quad g''=g_A
\end{eqnarray}
with $g_A=0.75$. A different set of coupling constants can be
found in Ref. \cite{behill}. With our notation, we get the
following relations
\begin{eqnarray}
g=g_A', \quad g'=-g_A', \quad h=\frac{1}{2}G_A
\end{eqnarray}
with $G_A\approx 1$ and $g_A'=0.6$ \cite{behill}. In fact, the
coupling constant $g_{A}$ was studied in many theoretical
approaches such as QCD sum rules \cite{QSR,QSR-1,QSR-2,QSR-3}. In
this work we use the experimental value $g_{A}=0.59\pm
0.07\pm0.01$ extracted from the width of $D^*$ \cite{isoda}. With
the available experimental information, Casalbuoni and
collaborators extracted $h=0.56\pm 0.28$ and
$h'=(h_1+h_2)/\Lambda_{\chi}=0.55$ GeV$^{-1}$ \cite{casalbuoni}.
The signs of $g,\;g'\;g''$ are not determined although their
absolute values are known.

The interaction Lagrangian related to the $\sigma$ meson can be
written as
\begin{eqnarray}
\mathcal{L}_\sigma&=&g_\sigma{\rm Tr}[H
\sigma\overline{H}]+g'_\sigma{\rm Tr}[S
\sigma\overline{S}]+g^{''}_\sigma{\rm
Tr}[T^\mu\sigma\overline{T}_\mu]\nonumber\\
&&+\frac{h_\sigma}{f_\pi}{\rm Tr}[S (\partial_\mu\sigma)
\gamma^\mu\overline{H}]+\frac{h'_\sigma}{f_\pi}{\rm Tr}[T^\mu
(\partial_\mu\sigma) \overline{H}]+h.c.\label{sig}
\end{eqnarray}
In order to estimate the values of the coupling constants, we
compare the above Lagrangian with that in Ref. \cite{behill} and
get
\begin{eqnarray}
g_\sigma=-\frac{1}{2\sqrt6}g_\pi,\quad
g_\sigma'=-\frac{1}{2\sqrt6}g_\pi, \quad
h_\sigma=\frac{1}{\sqrt3}g_A',
\end{eqnarray}
where $g_\pi=3.73$. When performing numerical analysis, we take
$g_\sigma''= g_\sigma$ and $h_\sigma'= h_\sigma$ approximately.

\section{Derivation of the pion and sigma exchange potential }
\label{sec3}

If $Z^{+}(4430)$ is a molecular state of $D_{1}'-D^*$ or
$D_{1}-D^*$, the flavor wavefunction of $Z^{+}(4430)$ reads
\cite{xiangliu},
\begin{eqnarray}
|Z^{+}(4430)\rangle=\frac{1}{\sqrt{2}}\Big [|\bar{D}_{1}^{'0}
D^{*+}\rangle+|\bar{D}^{*0}D_{1}^{'+}\rangle\Big]
\label{wave-1}\end{eqnarray} or
\begin{eqnarray}
|Z^{+}(4430)\rangle=\frac{1}{\sqrt{2}}\Big [|\bar{D}_{1}^0
D^{*+}\rangle+|\bar{D}^{*0}D_{1}^+\rangle\Big].\label{wave-2}
\end{eqnarray}
Here $D_{1}'$ and $D_1$ with quantum number $J^{P}=1^+$ belong to
$(0^+,1^+)$ and $(1^{+},2^+)$ doublet respectively in the heavy
quark limit.

To derive the effective potentials, we follow the same procedure
as in Ref. \cite{xiangliu}. Firstly we write out the elastic
scattering amplitudes of the direct process $A(B)\to A(B)$ and
crossed channel $A(B)\to B(A)$, where $A$ and $B$ denote
$D_1^{(')}$ and $D^*$.  Secondly, we impose the constraint that
initial states and final states should have the same angular
momentum. Thirdly, we average the potentials obtained with Breit
approximation in the momentum space. Finally we perform Fourier
transformation to derive the potentials in the coordinate space.
For the scattering between $D^\ast$ and $D_1'$ ($D_1$), both the
pion and sigma meson exchange are allowed in both direct and
crossed processes.

We introduce the form factor (FF) in every interaction vertex to
compensate the off-shell effects of the exchanged mesons when
writing out the scattering amplitude, which differs from Ref.
\cite{xiangliu}. One adopts the dipole type FF \cite{Tornqvist,FF}
\begin{eqnarray}
F(q)=\Big(\frac{\Lambda^2-m^2}{\Lambda^2-q^2}\Big)^2 \; .
\end{eqnarray}
The phenomenological parameter $\Lambda$ is near 1 GeV. $q$
denotes the four-momentum of the exchange meson. It is observed
that as $q^2\to 0$ it becomes a constant and if $\Lambda\gg M$, it
turns to be unity. In the case, as the distance is infinitely
large, the vertex looks like a perfect point, so the form factor
is simply 1 or a constant. Whereas, as $q^2\rightarrow \infty$,
the form factor approaches to zero, namely, in this situation, the
distance becomes very small, the inner structure (quark, gluon
degrees of freedom) would manifest itself and the whole picture of
hadron interaction is no longer valid, so the form factor is zero
which cuts off the end effects \cite{FF}.

For the direct scattering channel, $q_0$ is a small value. In the
heavy quark limit we approximately take $q_0=0$. Thus it is
reasonable to approximate $q^2$ as $-\mathbf{q}^2$. However, for
the crossed diagram, we can not ignore the contribution from $q_0$
due to $q_0=M_{D_1}-M_{D^*}\approx 410$ MeV, which is about three
times of the pion mass. The principal integration is a good
approach to solve the problem when $q_0$ is larger than the mass
of exchanging meson.

Since we only consider S-wave bound states between $D_1^{'}(D_1)$
and $D^*$, there are five independent parts related to the
potentials of $D_1^{'}(D_1)-D^*$ system in the momentum space. We
use the following definitions to denote them after performing
Fourier transformation.
\begin{eqnarray}
\frac{1}{\mathbf{q}^2+m^2}&\rightarrow& Y_0(\Lambda, m, r),\\
\frac{\mathbf{q}^2}{\mathbf{q}^2+m^2}&\rightarrow& Y_1(\Lambda, m, r),\\
\frac{q_0^2}{q^2-m^2}&\rightarrow& Y_2(\Lambda, m, r),\\
\frac{(\mathbf{q}^2)^2}{q^2-m^2}&\rightarrow& Y_3(\Lambda, m, r),\\
\frac{\mathbf{q}^2}{q^2-m^2}&\rightarrow& Y_4(\Lambda, m, r),
\end{eqnarray}
where $m$ denotes the mass of the $\pi$ or $\sigma$ meson. Their
explicit expressions are
\begin{eqnarray}
Y_0(\Lambda, m_\sigma, r)&=&\frac{1}{4\pi r}(e^{-m_\sigma
r}-e^{-\Lambda r})-\frac{\eta'^2}{8\pi\Lambda}e^{-\Lambda
r}-\eta'^4\xi(\Lambda)-\eta'^6\zeta(\Lambda),\\
Y_1(\Lambda, m_\pi, r)&=&-\frac{m_\pi^2}{4\pi r}(e^{-m_\pi
r}-e^{-\Lambda
r})+\frac{m_{\pi}^2\eta^2}{8\pi\Lambda}e^{-\Lambda r}+m_{\pi}^2\eta^4\xi(\Lambda)+\Lambda^2\eta^6\zeta(\Lambda),\\
Y_2(\Lambda, m_\pi, r)&=&-\frac{q_0^2}{4\pi r}[\cos(\mu
r)-e^{-\alpha r}]+\frac{q_0^2\eta^2}{8\pi\alpha}e^{-\alpha r}
+q_0^2\eta^4\xi(\alpha)+q_0^2\eta^6\zeta(\alpha),\\
Y_3(\Lambda, m_\pi, r)&=&-\frac{\mu^4}{4\pi r}\cos(\mu
r)+\frac{\mu^4}{4\pi r}e^{-\alpha
r}+\frac{\eta^2\mu^4}{8\pi\alpha}e^{-\alpha r}
+\eta^4 (-\alpha^4-2\mu^2\alpha^2)\xi(\alpha)+\alpha^4\eta^6\zeta(\alpha),\\
Y_4(\Lambda, m_\sigma, r)&=&\frac{\mu'^2}{4\pi r}[e^{-\mu'
r}-e^{-\alpha r}]-\frac{\eta'^2\mu'^2}{8\pi\alpha}e^{-\alpha
r}-\mu'^2\eta'^4\xi(\alpha)-\alpha^2\eta'^6\zeta(\alpha)
\end{eqnarray}
with
\begin{eqnarray}
\xi(a)&=&\frac{e^{-a r}r}{32\pi a^2} +\frac{e^{- a r}}{32\pi
a^3},\\
\zeta(b)&=&\frac{e^{-b r}r^2}{192\pi b^3} +\frac{e^{-b r}r}{64\pi
b^4}+\frac{e^{-b r}}{64\pi b^5},
\end{eqnarray}
where $\mu=\sqrt{q_0^2-m_\pi^2}$, $\mu'=\sqrt{m_\sigma^2-q_0^2}$,
$\eta=\sqrt{\Lambda^2-m_\pi^2}$,
$\eta'=\sqrt{\Lambda^2-m_\sigma^2}$ and
$\alpha=\sqrt{\Lambda^2-q_0^2}$. With the above functions, the
potentials for the different cases are collected in Table
\ref{table-1} and \ref{table-2}.

\begin{center}
\begin{table}[htb]
\begin{tabular}{c||cc|cc}\hline
&\multicolumn{2}{c}{$V_{\mathrm{Dir}}(r)$}\vline&\multicolumn{2}{c}{$V_{\mathrm{Cro}}(r)$}\\\hline\hline
&$\pi$ exchange&$\sigma$ exchange&$\pi$ exchange&$\sigma$ exchange \\
J=0& $\frac{gg'}{3f_\pi^2}Y_1(\Lambda,m_\pi,r)$&$g_\sigma g_\sigma' Y_0(\Lambda,m_\sigma,r)$&$-\frac{h^2}{2f_\pi^2}Y_2(\Lambda,m_\pi,r)$&$\frac23\frac{h_\sigma^2}{f_\pi^2}Y_4(\Lambda,m_\sigma,r)$\\
J=1&
$\frac{gg'}{6f_\pi^2}Y_1(\Lambda,m_\pi,r)$&$g_\sigma g_\sigma' Y_0(\Lambda,m_\sigma,r)$&$-\frac{h^2}{2f_\pi^2}Y_2(\Lambda,m_\pi,r)$&$\frac13\frac{h_\sigma^2}{f_\pi^2}Y_4(\Lambda,m_\sigma,r)$\\
J=2& $-\frac{gg'}{6f_\pi^2}Y_1(\Lambda,m_\pi,r)$&$g_\sigma
g_\sigma'
Y_0(\Lambda,m_\sigma,r)$&$-\frac{h^2}{2f_\pi^2}Y_2(\Lambda,m_\pi,r)$&$-\frac13\frac{h_\sigma^2}{f_\pi^2}Y_4(\Lambda,m_\sigma,r)$\\\hline
\end{tabular}
\caption{The potentials with the $D_1'$-$D^\ast$ molecular
assumption for $Z^+(4430)$.\label{table-1}}
\end{table}

\begin{table}[htb]
\begin{tabular}{c||cc|cc}\hline
&\multicolumn{2}{c}{$V_{\mathrm{Dir}}(r)$}\vline&\multicolumn{2}{c}{$V_{\mathrm{Cro}}(r)$}\\\hline\hline
&$\pi$ exchange&$\sigma$ exchange&$\pi$ exchange&$\sigma$ exchange\\
J=0& $-\frac{5gg''}{18f_\pi^2}Y_1(\Lambda,m_\pi,r)$&$g_\sigma g_\sigma'' Y_0(\Lambda,m_\sigma,r)$&$\frac{(h')^2}{6f_\pi^2}Y_3(\Lambda,m_\pi,r)$&$\frac19\frac{(h'_\sigma)^2}{f_\pi^2}Y_4(\Lambda,m_\sigma,r)$\\
J=1&
$-\frac{5gg''}{36f_\pi^2}Y_1(\Lambda,m_\pi,r)$&$g_\sigma g_\sigma'' Y_0(\Lambda,m_\sigma,r)$&$-\frac{(h')^2}{12f_\pi^2}Y_3(\Lambda,m_\pi,r)$&$\frac{1}{18}\frac{(h'_\sigma)^2}{f_\pi^2}Y_4(\Lambda,m_\sigma,r)$\\
J=2& $\frac{5gg''}{36f_\pi^2}Y_1(\Lambda,m_\pi,r)$&$g_\sigma
g_\sigma''
Y_0(\Lambda,m_\sigma,r)$&$\frac{(h')^2}{60f_\pi^2}Y_3(\Lambda,m_\pi,r)$&$-\frac{1}{18}\frac{(h'_\sigma)^2}{f_\pi^2}Y_4(\Lambda,m_\sigma,r)$\\\hline
\end{tabular}
\caption{The effective potentials with the assumption of
$Z^+(4430)$ being a $D_1$-$D^\ast$ molecule. \label{table-2}}
\end{table}
\end{center}

Assuming $Z^+(4430)$ to be $D_1'D^*$ or $D_1D^*$ molecule state,
the total potential is
\begin{eqnarray}
V_{\mathrm{Total}}(r)=V_{\mathrm{Dir}}(r)+V_{\mathrm{Cro}}(r),
\end{eqnarray}
where the sign between $V_{\mathrm{Dir}}(r)$ and
$V_{\mathrm{Cro}}(r)$ is determined by the flavor wavefunction of
$Z^+(4430)$ in Eq. (\ref{wave-1}) or (\ref{wave-2}).
$V_{\mathrm{Dir}}(r)$ and $V_{\mathrm{Cro}}(r)$ correspond to the
potentials from the direct and crossed diagram respectively. From
Tables \ref{table-1} and \ref{table-2} we make two interesting
observations: (1)$V_{\mathrm{Cro}}(r)$ does not depend on the sign
of coupling constant; (2)For the $D_1'D^*$ systems, the sigma
exchange potentials for the direct diagram and the pion exchange
potentials for the crossed diagram are the same for the three
cases with $J=0,1,2$!

\section{The shape of the pion and sigma exchange potential }
\label{sec4}

In this section we study the variation of the pion and sigma meson
exchange potentials with the coupling constants, which are given
in the Section \ref{sec2}. We also need the following input
parameters $m_{D^*}=2008.35$ MeV, $m_{D_1'}=2427$ MeV,
$m_{D_1}=2422.85$ MeV; $f_{\pi}=132$ MeV, $m_{\pi}=135.0$ MeV,
$m_{\sigma}=600$ MeV \cite{PDG}.

\subsection{Single pion exchange potential}

\begin{figure}[htb]
\begin{center}
\begin{tabular}{ccc}
\scalebox{0.5}{\includegraphics{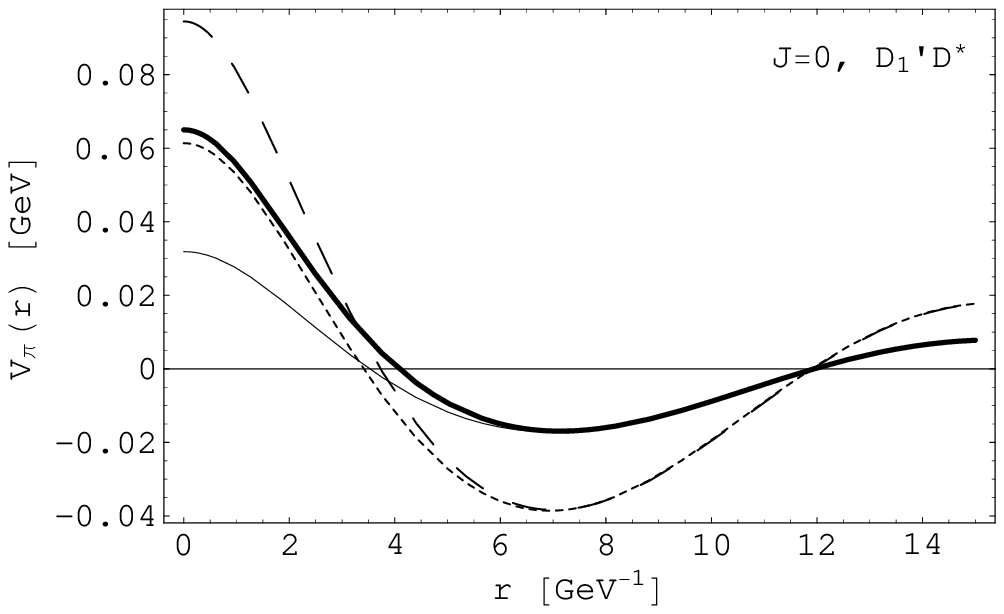}}&
\scalebox{0.5}{\includegraphics{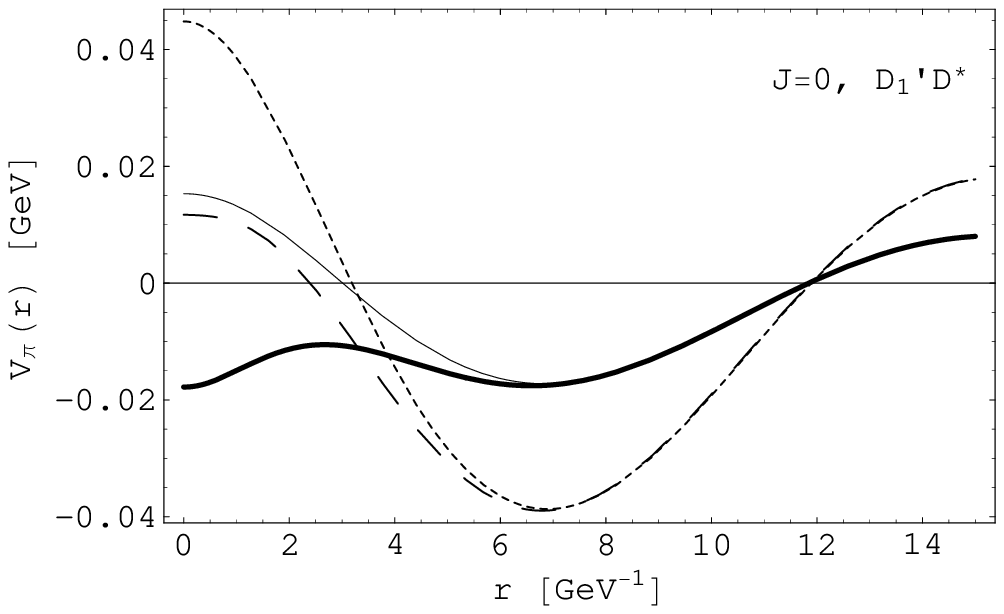}}&
\scalebox{0.5}{\includegraphics{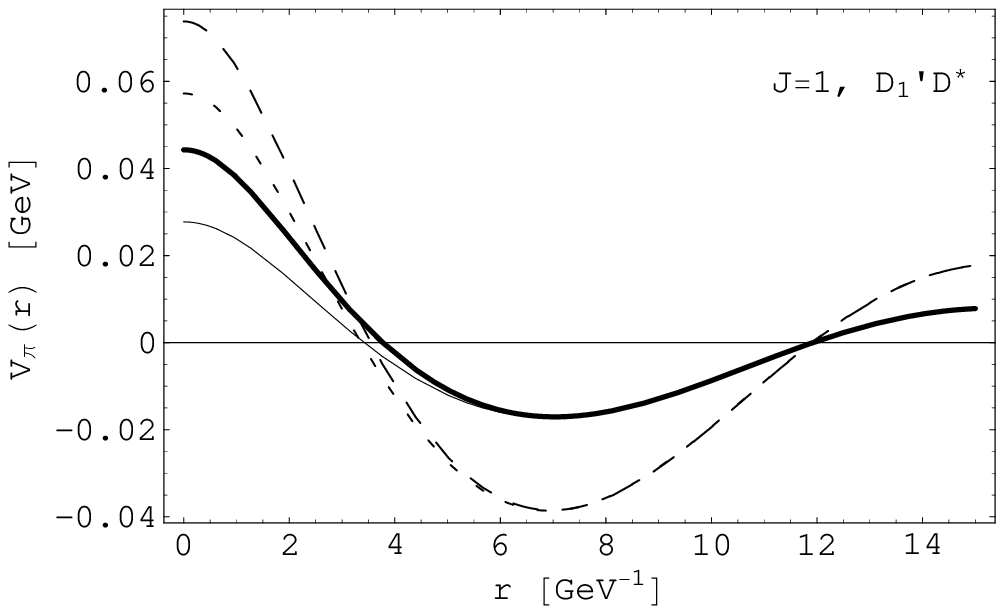}} \\
(a)&(b)&(c)\\
\scalebox{0.5}{\includegraphics{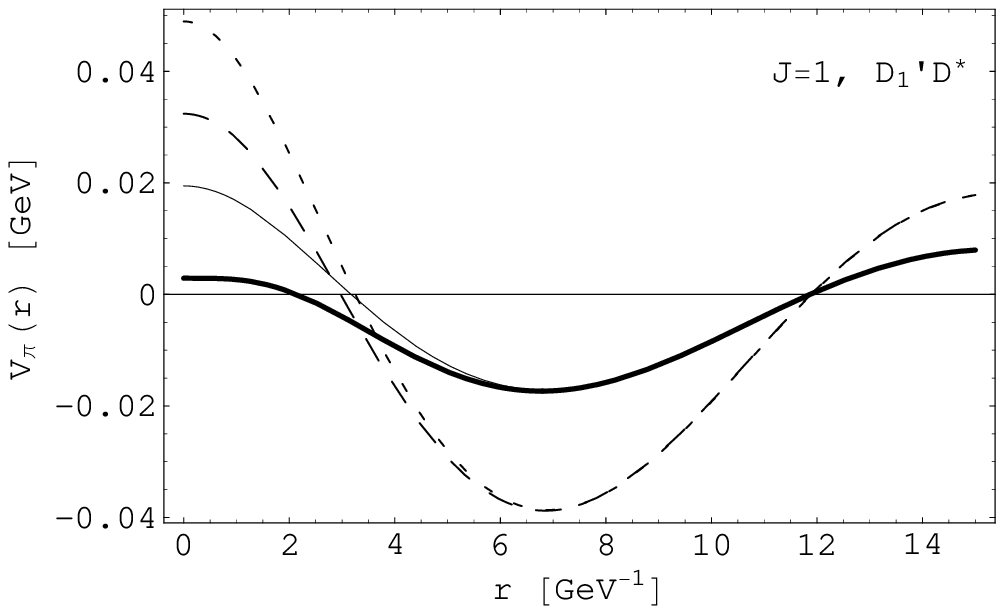}}&
\scalebox{0.5}{\includegraphics{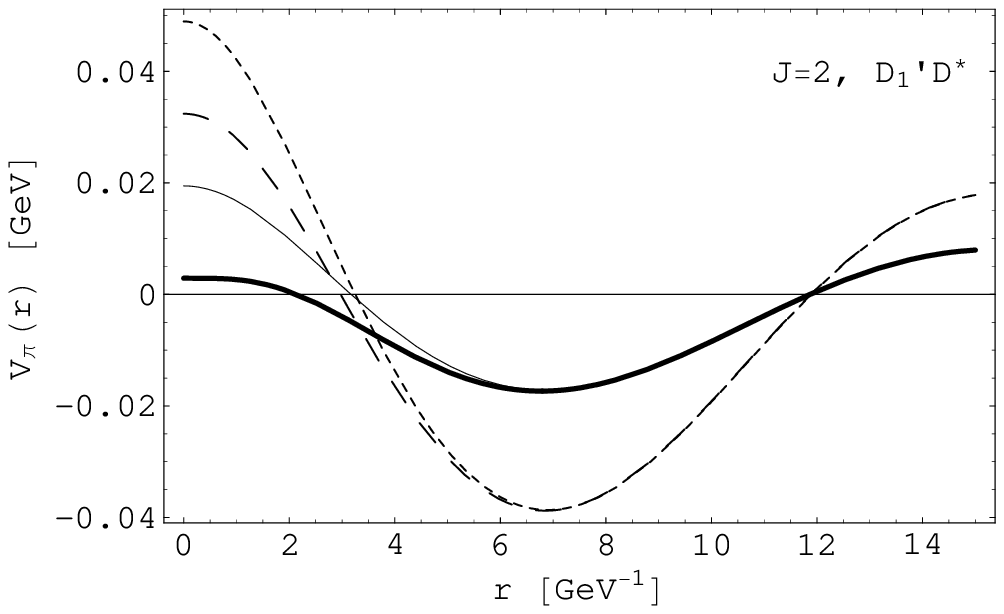}}&
\scalebox{0.5}{\includegraphics{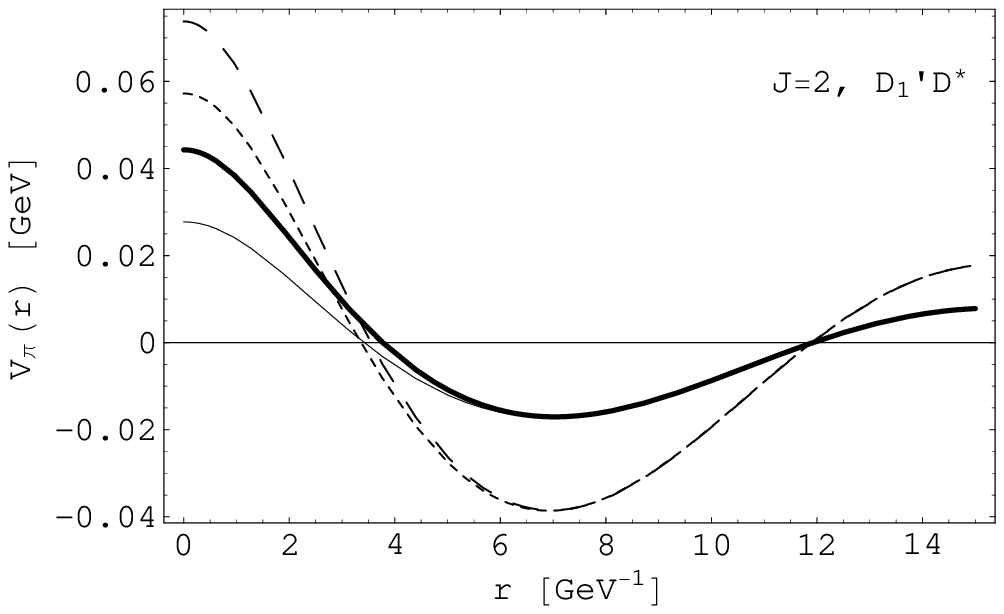}}\\
(d)&(e)&(f)\\
\end{tabular}
\caption{The single pion exchange potential for the $D_1'D^*$
molecule. In diagrams (a), (c) and (e), the solid, thick solid,
dotted and dashed lines correspond to the potentials with
parameters [$g\cdot g'$, $h$]=[0.1, 0.56], [0.5, 0.56], [0.1,
0.84], [0.5, 0.84] respectively. In (b), (d) and (f), the solid ,
thick solid, dotted and dashed lines correspond to the potentials
with parameters [$g\cdot g'$, $h$]=[-0.1,0.56], [-0.5, 0.56],
[-0.1, 0.84], [-0.5, 0.84] respectively. Here $\Lambda=1$ GeV.
\label{pp-1}}
\end{center}
\end{figure}

\begin{center}
\begin{figure}[htb]
\begin{tabular}{ccc}
\scalebox{0.5}{\includegraphics{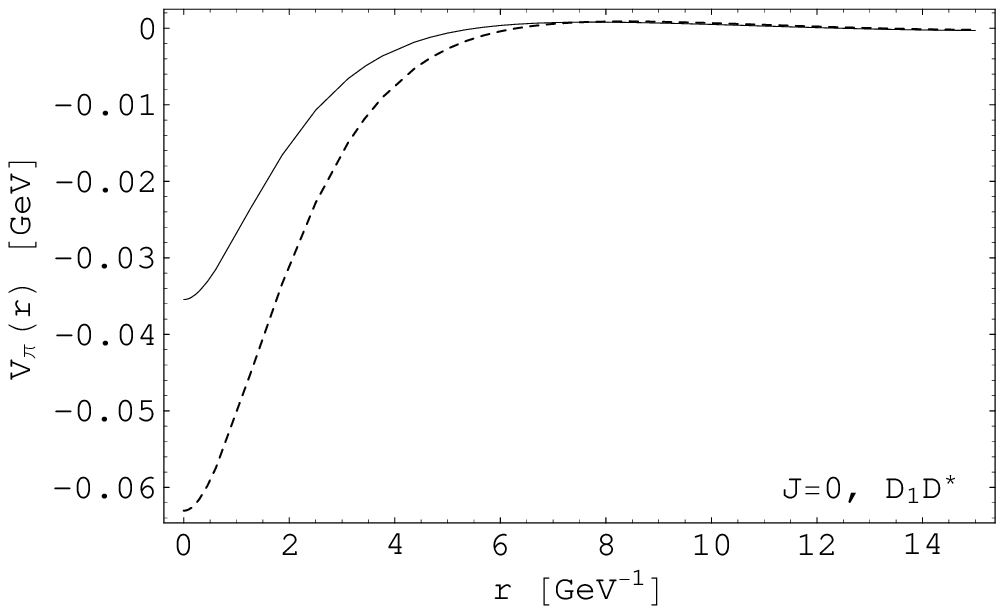}}&
\scalebox{0.5}{\includegraphics{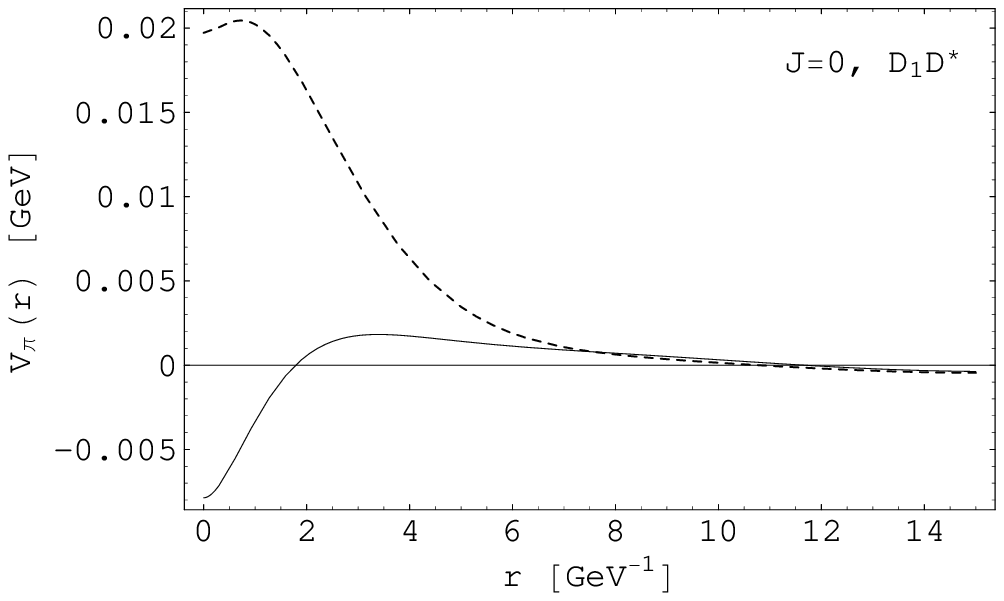}}&
\scalebox{0.5}{\includegraphics{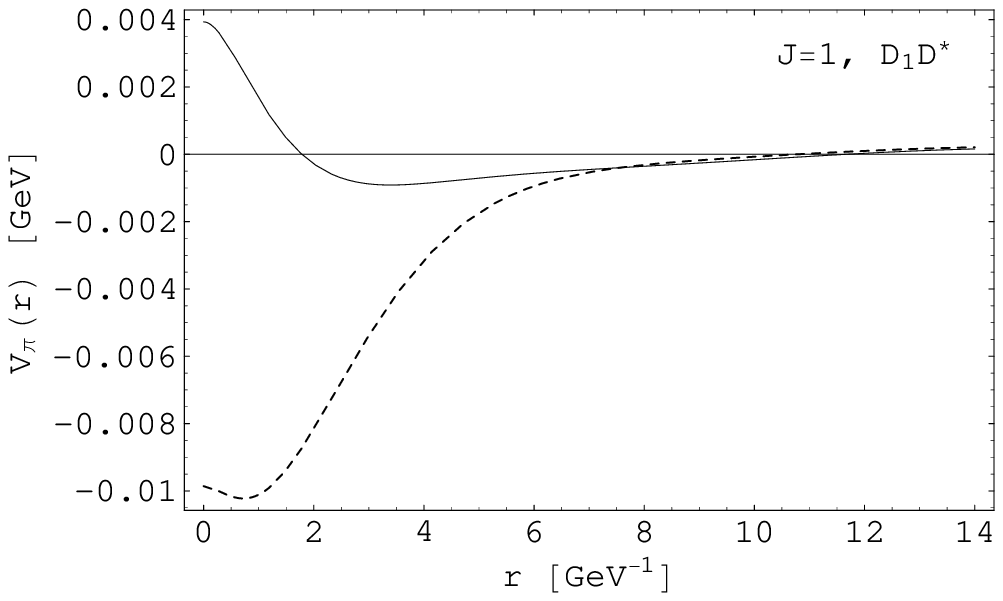}} \\
(a)&(b)&(c)\\
\scalebox{0.5}{\includegraphics{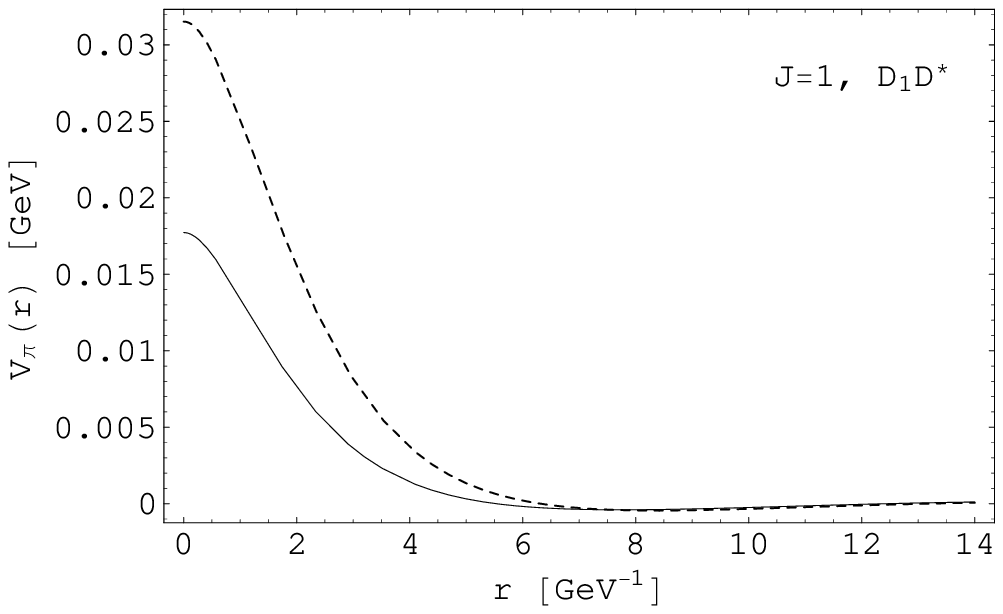}}&
\scalebox{0.5}{\includegraphics{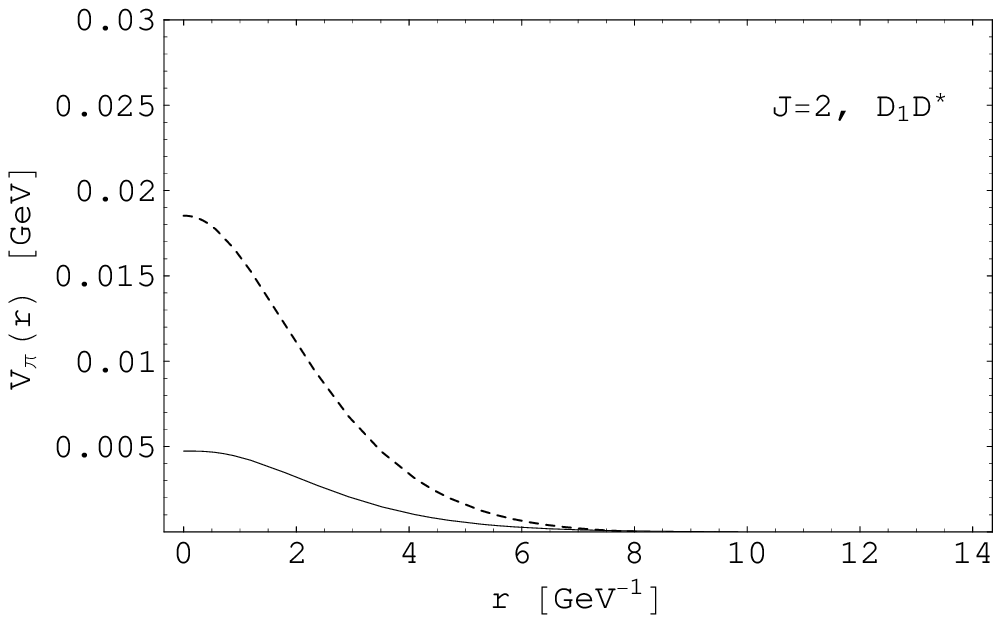}}&
\scalebox{0.5}{\includegraphics{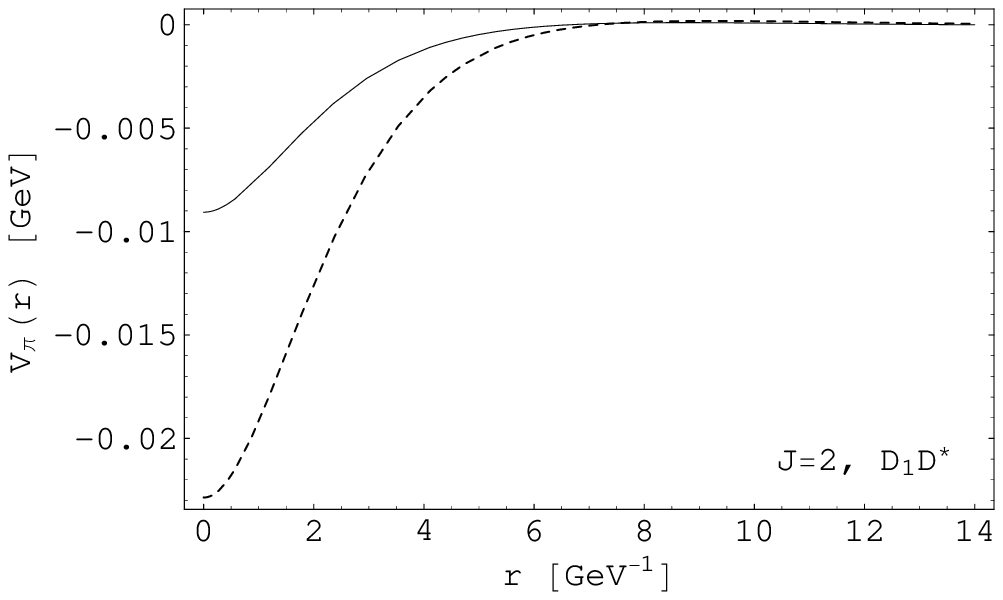}} \\
(d)&(e)&(f)\\
\end{tabular}
\caption{The single pion exchange potential for the $D_1D^*$
molecule. In (a), (c), (e), the solid and dotted lines denote the
potentials with parameters [$g\cdot g''$, $h'$]=[0.2, 0.55
GeV$^{-1}$], [0.6, 0.55 GeV$^{-1}$] respectively. In (b), (d),
(f), the solid and dotted line denote the potentials with
parameters [$g\cdot g'', h'$]=[-0.2, 0.55 GeV$^{-1}$], [-0.6, 0.55
GeV$^{-1}$] respectively. Here $\Lambda=1$ GeV.\label{pp-2}}
\end{figure}
\end{center}

For the S-wave $D_{1}'D^*$ system, we take several typical values
of the coupling constants $g\cdot g'=\pm0.1, \pm0.5$ and $h=0.56,
0.84$. Meanwhile we take the cutoff $\Lambda=1$ GeV. We illustrate
the dependence of single pion exchange potential on these typical
values in Fig. \ref{pp-1}.  We also plot the potential of the
$D_{1}D^*$ molecular state with several typical coupling constants
$[g\cdot g''$, $h'$]=[$\pm0.2$, 0.55 GeV$^{-1}$], [$\pm0.6$, 0.55
GeV$^{-1}$] in Fig. \ref{pp-2}.


From Fig. \ref{pp-1} and \ref{pp-2}, we notice that (1) the
variation of $g\cdot g'$ or $g\cdot g''$ does not result in the
big change of the potential when $r$ is larger than 6 GeV$^{-1}$;
(2) the potential is sensitive to the value of $h$, which
indicates the single pion exchange in the crossed diagram plays an
important role to bind the $D_{1}^{'}D^*$ compared with the
contribution of direct diagram; (3) for the S-wave $D_{1}D^*$
system with $J=0$, its potential is repulsive with $[g\cdot g'',
h']$=[-0.6, 0.55 GeV$^{-1}$]. The potentials of the $J=1$
$D_{1}D^*$ system with [$g\cdot g'', h'$]=[-0.2, 0.55 GeV$^{-1}$],
[-0.6, 0.55 GeV$^{-1}$] and the $J=2$ $D_{1}D^*$ system with
[$g\cdot g''$, $h'$]=[0.2, 0.55 GeV$^{-1}$], [0.6, 0.55
GeV$^{-1}$] are also repulsive, which are shown in Fig. \ref{pp-2}
(b), (d) and (e). In the range $r<6$ GeV$^{-1}$, the potential of
the $D_1-D^*$ system is sensitive to the coupling constants. These
conclusions were obtained when taking $\Lambda=1$ GeV.

\subsection{The potential via both pion and sigma exchanges}

\begin{figure}[htb]
\begin{center}
\begin{tabular}{cccc}
\scalebox{0.42}{\includegraphics{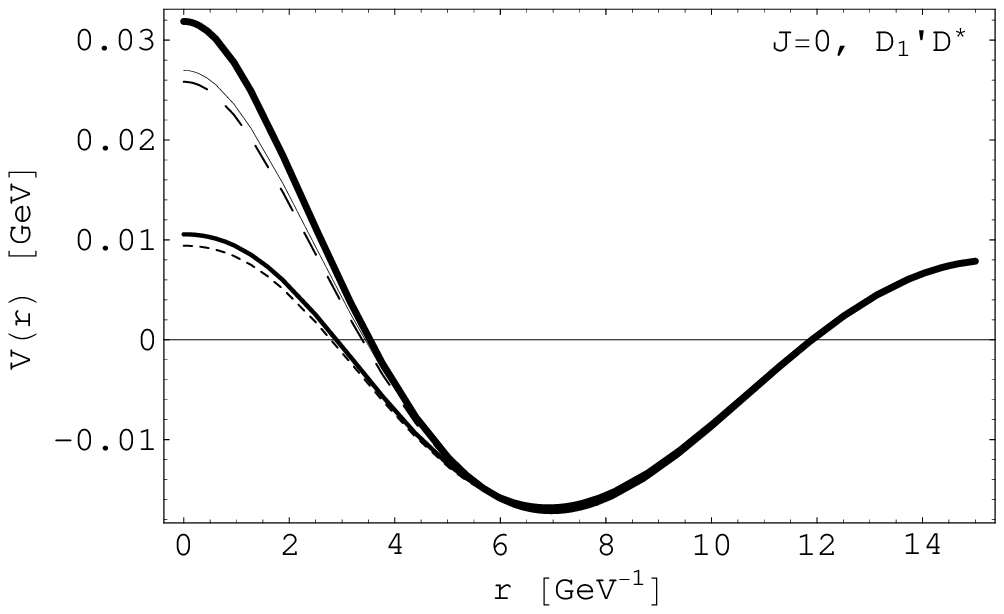}}&
\scalebox{0.42}{\includegraphics{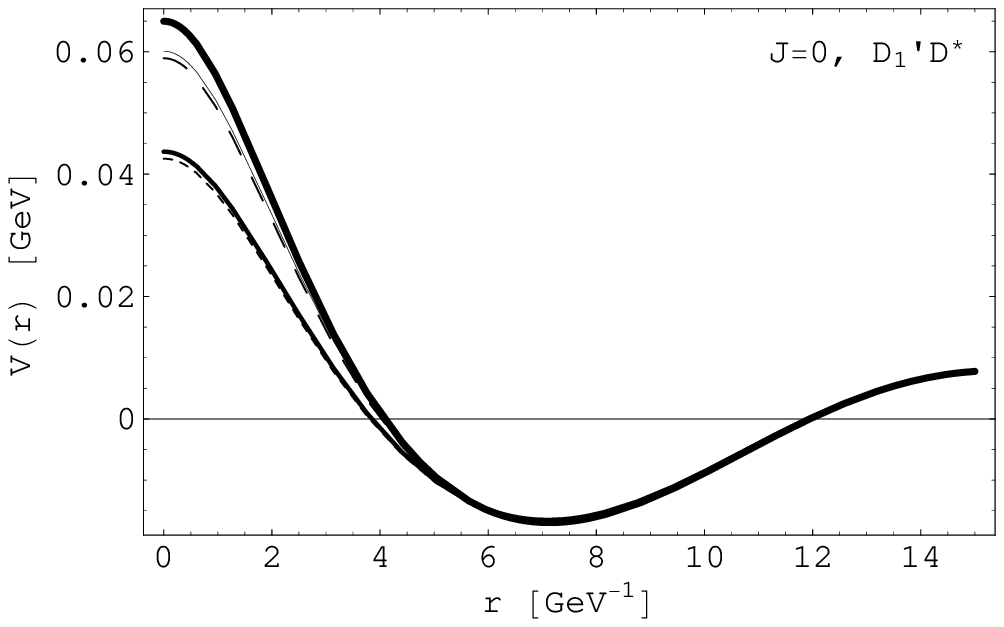}}&
\scalebox{0.42}{\includegraphics{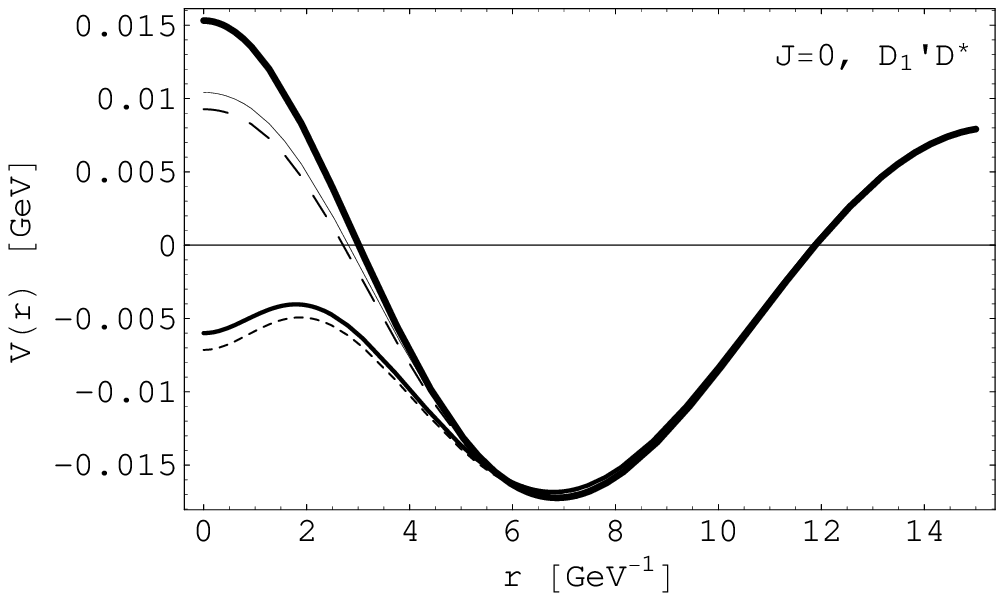}}&
\scalebox{0.42}{\includegraphics{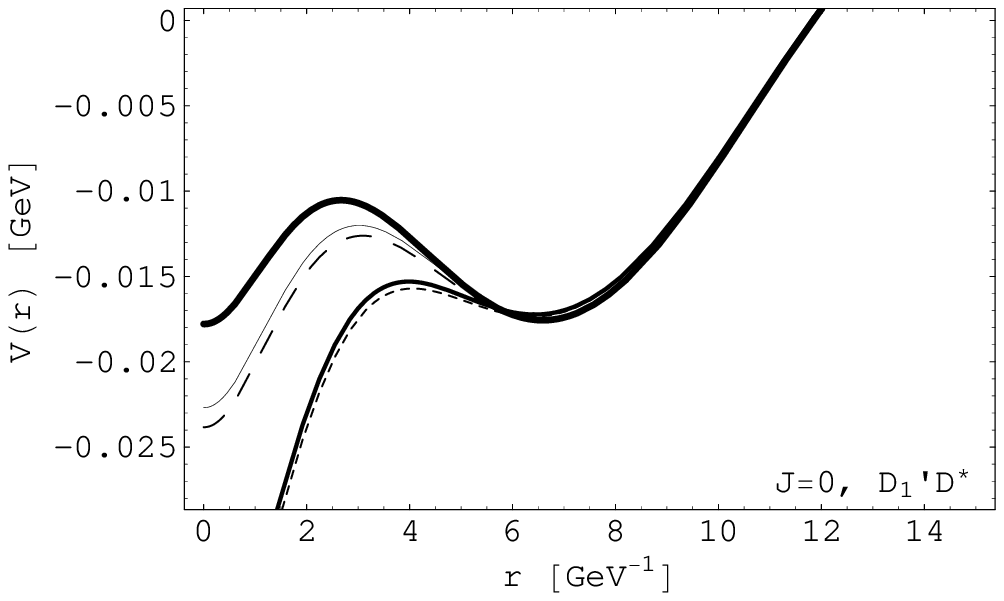}}\\
(a-1)&(a-2)&
(a-3)&(a-4)\\
\end{tabular}
\caption{For the $D_1'D^*$ molecular state with $J=0$, we compare
the single pion exchange potential with the total potentials
containing the sigma exchange. The thick solid lines denote the
single pion exchange potentials. The solid lines, thin solid line,
dashed line and dotted line correspond to the potentials with
parameters [$g_{\sigma}g_{\sigma}', h_{\sigma}$]=[0.58, 0.4],
[0.58, 0.8], [-0.58, 0.4], [-0.58, 0.8] respectively. (a-1),
(a-2), (a-3) and (a-4) respectively correspond to [$g \cdot g'$ =
0.1, $h$ = 0.56], [$g\cdot g'$ = 0.5, $h$ = 0.56], [$g\cdot g'$ =
0.1, $h$ = 0.84] and [$g\cdot g'$ = 0.5, $h$ = 0.84]. Here
$\Lambda=1$ GeV. \label{pp-3}}

\begin{tabular}{cccc}
\scalebox{0.42}{\includegraphics{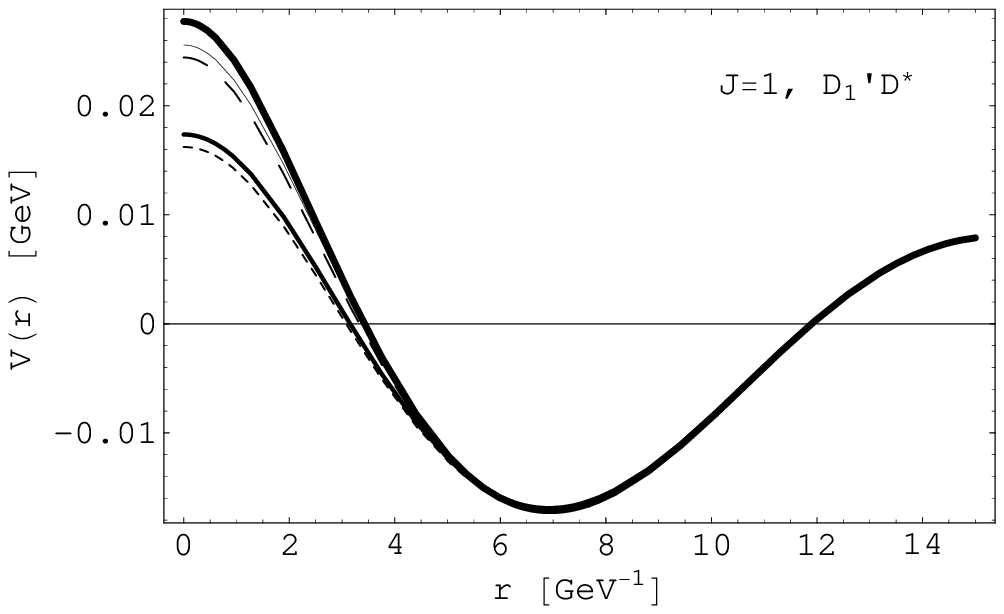}}&
\scalebox{0.42}{\includegraphics{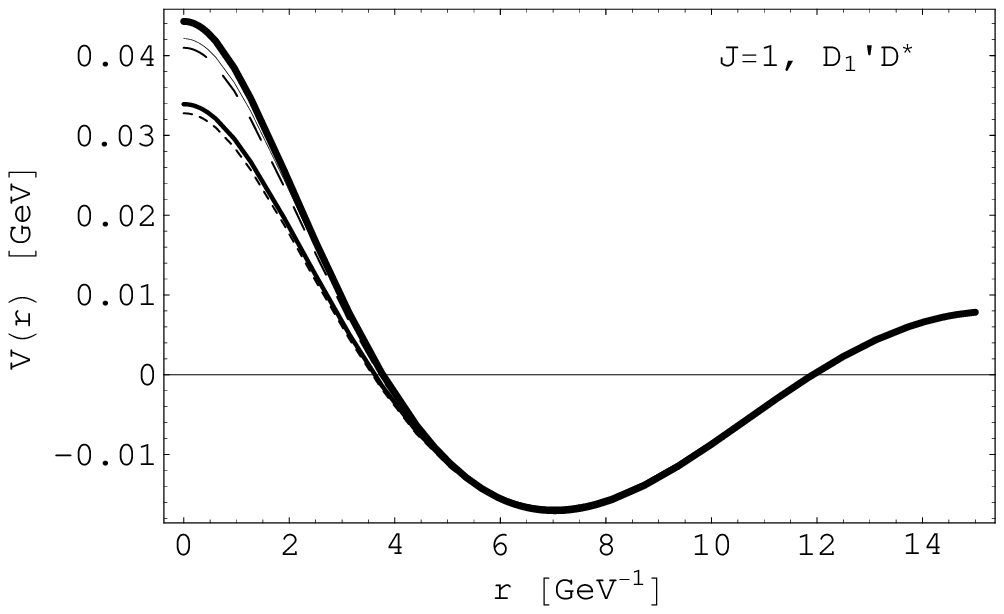}}&
\scalebox{0.42}{\includegraphics{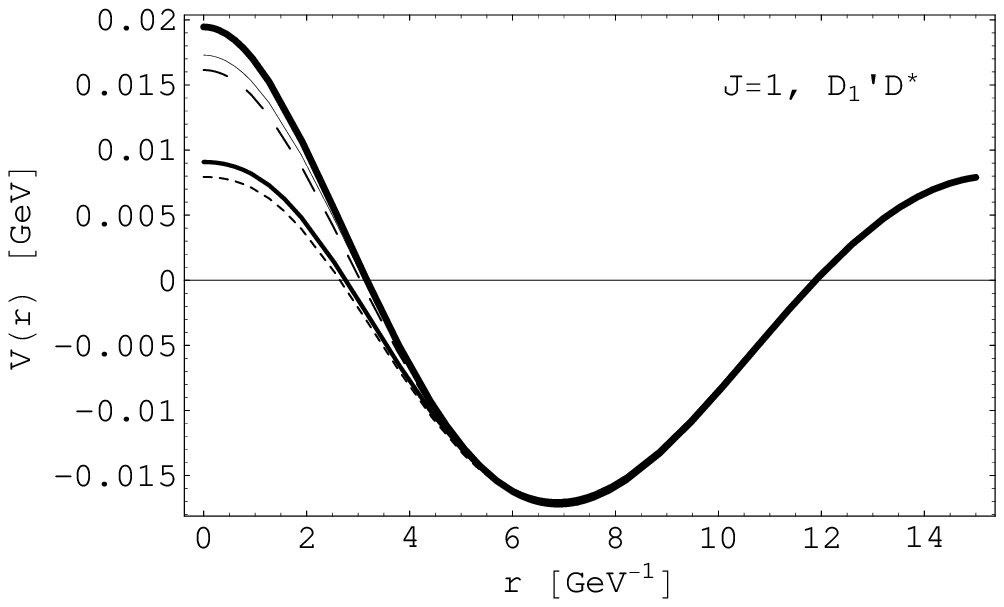}}&
\scalebox{0.42}{\includegraphics{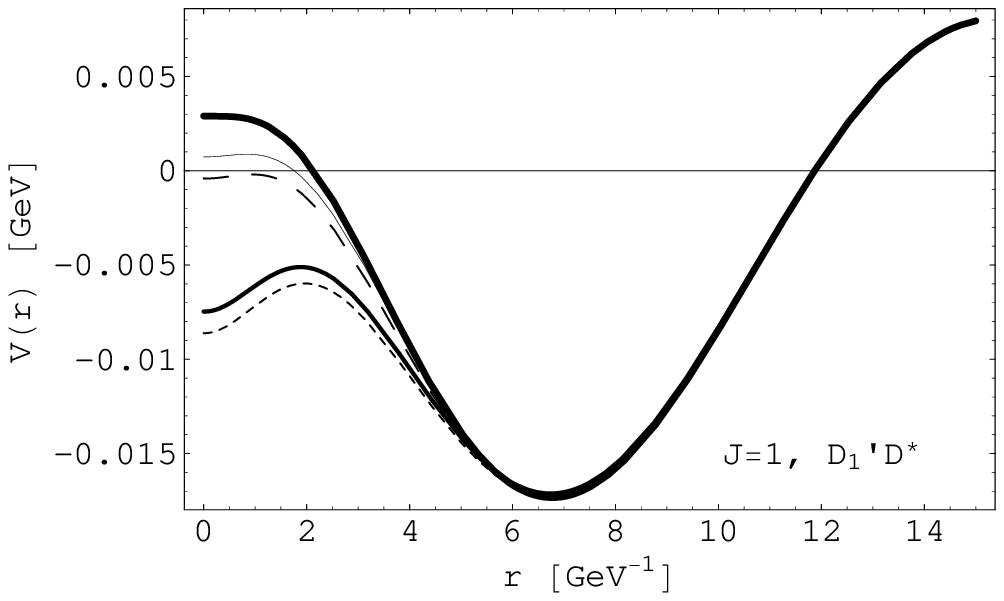}}\\
(a-1)&(a-2)&(a-3)&(a-4)\\
\end{tabular}
\caption{For the $D_1'D^*$ molecular state with $J=1$, we compare
the single pion exchange potential with the total potentials. The
thick solid lines denote the single pion exchange potentials. The
solid lines, thin solid line, dashed line and dotted line
correspond to the potentials with parameters
[$g_{\sigma}g_{\sigma}'$, $h_{\sigma}$]=[0.58, 0.4], [0.58, 0.8],
[-0.58, 0.4], [-0.58, 0.8] respectively. (a-1), (a-2), (a-3) and
(a-4) with [$g\cdot g'$ = 0.1, $h$ = 0.56], [$g\cdot g'$ = 0.5,
$h$ = 0.56], [$g\cdot g'$ = 0.1, $h$ = 0.84] and [$g\cdot g'$ =
0.5, $h$ = 0.84]. Here $\Lambda=1$ GeV.\label{pp-4}}
\end{center}
\end{figure}

\begin{figure}[htb]
\begin{center}
\begin{tabular}{cccc}
\scalebox{0.42}{\includegraphics{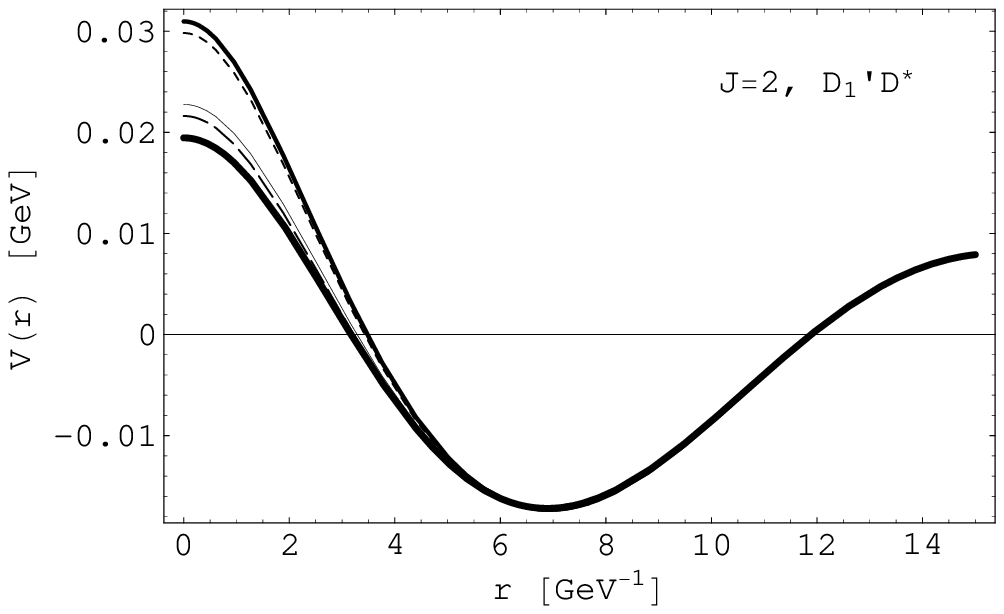}}&
\scalebox{0.42}{\includegraphics{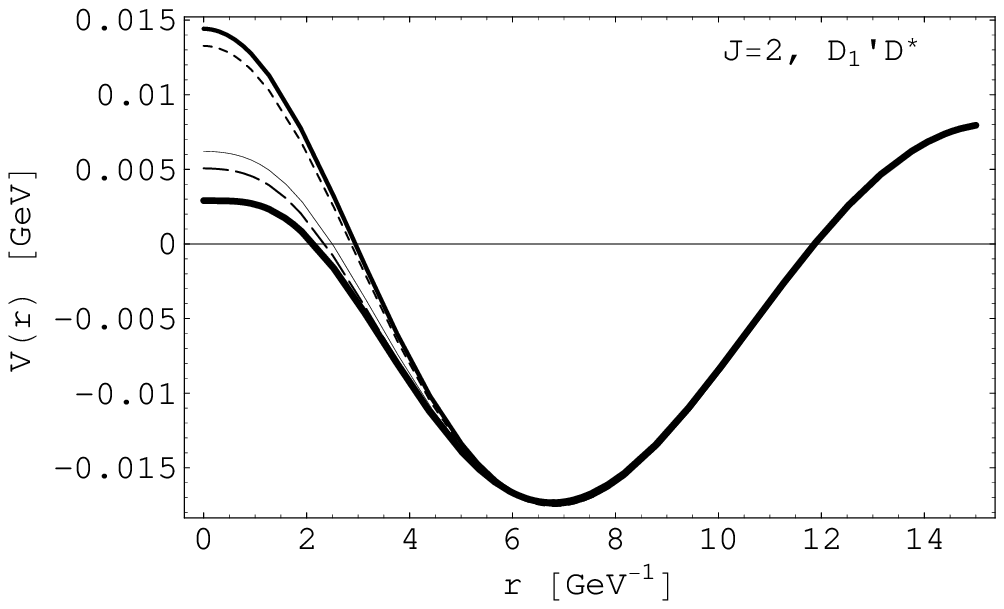}}&
\scalebox{0.42}{\includegraphics{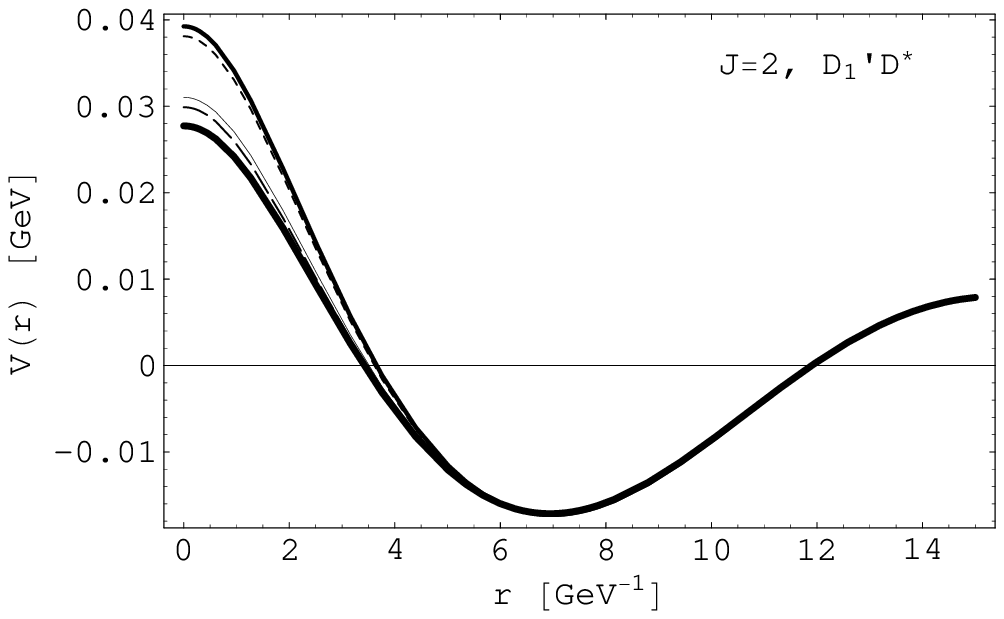}}&
\scalebox{0.42}{\includegraphics{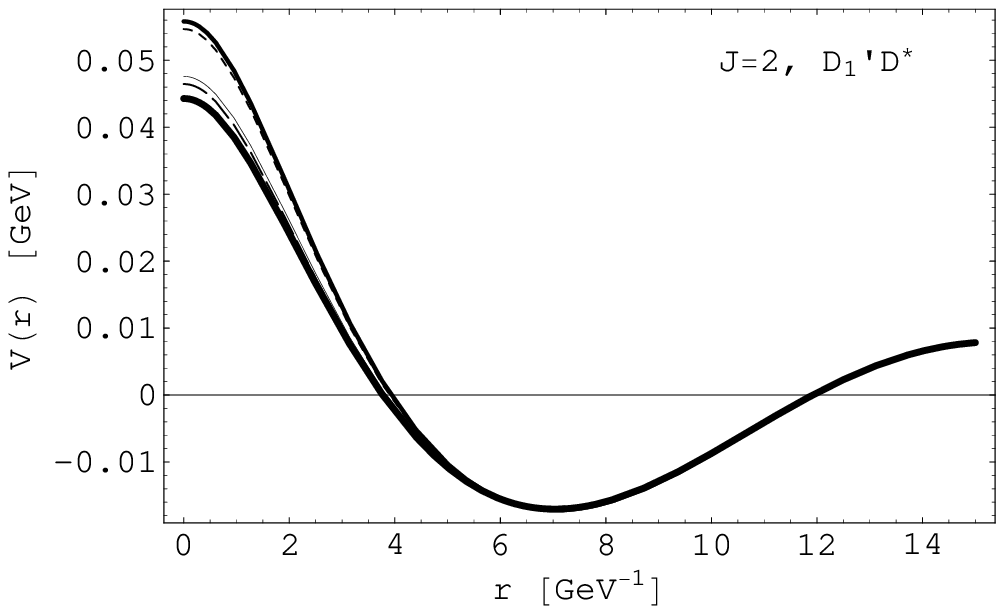}}\\
(a-1)&(a-2)&
(a-3)&(a-4)\\
\end{tabular}
\caption{For the $D_1'D^*$ molecular state with $J=2$, we compare
the single pion exchange potential with the total potentials. The
thick solid lines denote the single pion exchange potentials. The
solid lines, thin solid line, dashed line and dotted line
correspond to the potentials with parameters
[$g_{\sigma}g_{\sigma}'$, $h_{\sigma}$]=[0.58,0.4], [0.58,0.8],
[-0.58,0.4], [-0.58,0.8] respectively. (a-1), (a-2), (a-3) and
(a-4) with [$g\cdot g'$ = 0.1, $h$ = 0.56], [$g\cdot g'$ = 0.5, h
= 0.56], [$g\cdot g'$ = 0.1, $h$ = 0.84] and [$g\cdot g'$ = 0.5,
$h$ = 0.84].
 Here $\Lambda=1$ GeV.\label{pp-5}}
\end{center}
\end{figure}

We further investigate the potential of the $D_1'D^*$ system by
adding sigma exchange contribution. The typical values of coupling
constants are given in the captions of Figs.
\ref{pp-3}-\ref{pp-5}. In these figures, we compare the single
pion exchange potential with the total potential with different
coupling constants.

From Figs. \ref{pp-3}-\ref{pp-5}, one finds that adding sigma
exchange contribution does not result in dramatic change of the
total potential $V_{\mathrm{Total}}(r)$ of the S-wave $D_1'D^*$
and $D_1D^*$ molecule system especially when $r$ is larger than 6
GeV$^{-1}$.

\section{Numerical results}\label{sec5}

Different from the analysis in Ref. \cite{xiangliu}, in this work
we solve the the Schr\"{o}dinger equation numerically with the
help of MATSLISE package, which is a graphical Matlab software
package for the numerical study of regular Sturm-Liouville
problems, one-dimensional Schr\"{o}dinger equations and radial
Schr\"{o}dinger equations with a distorted coulomb potential. It
allows the fast and accurate computation of the eigenvalues and
the visualization of the corresponding eigenfunctions
\cite{matslise}.

\subsection{S-wave $D_{1}'-D^*$ system}

\begin{figure}[htb]
\begin{center}
\begin{tabular}{cc}
\scalebox{0.4}{\includegraphics{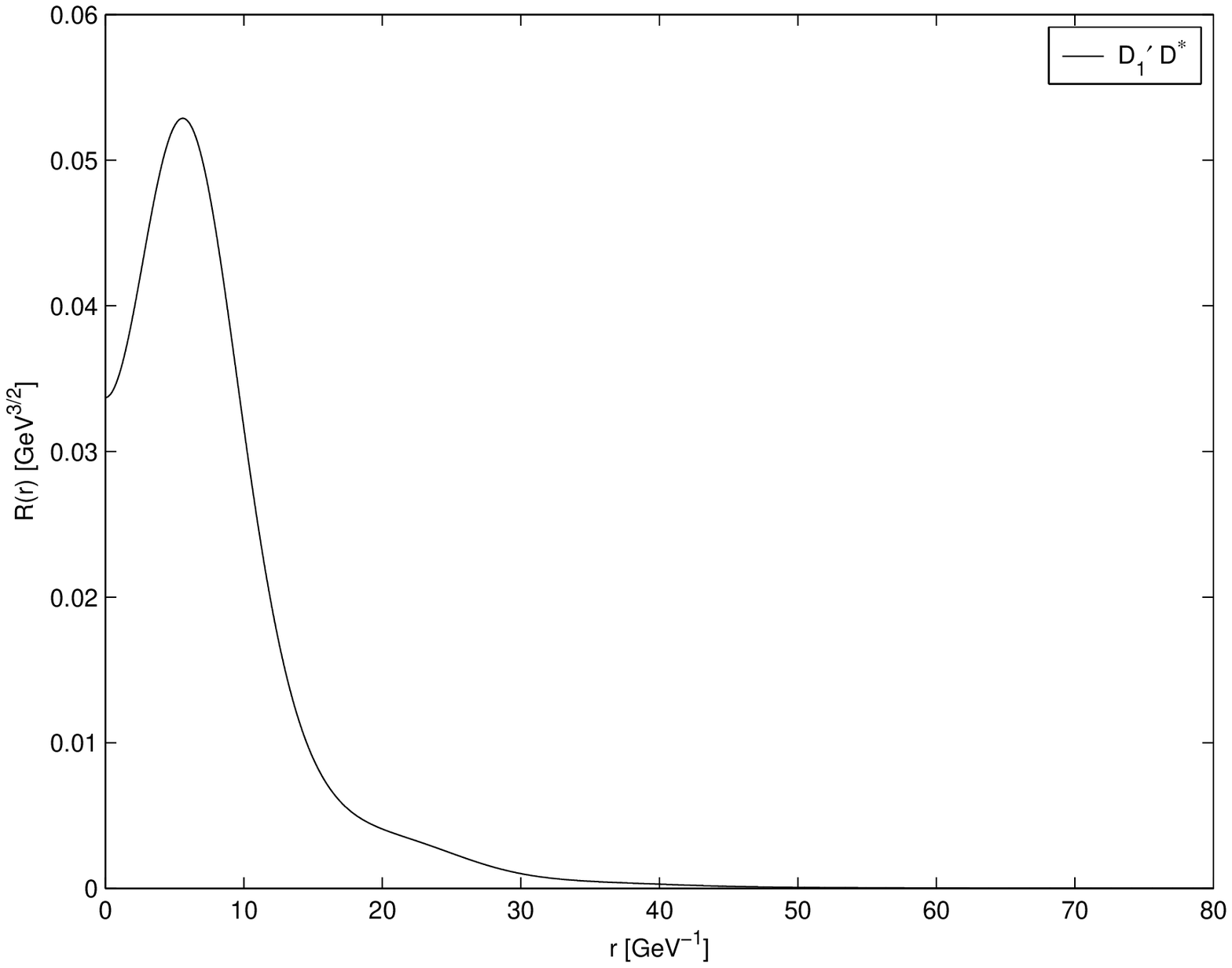}}&\scalebox{0.4}{\includegraphics{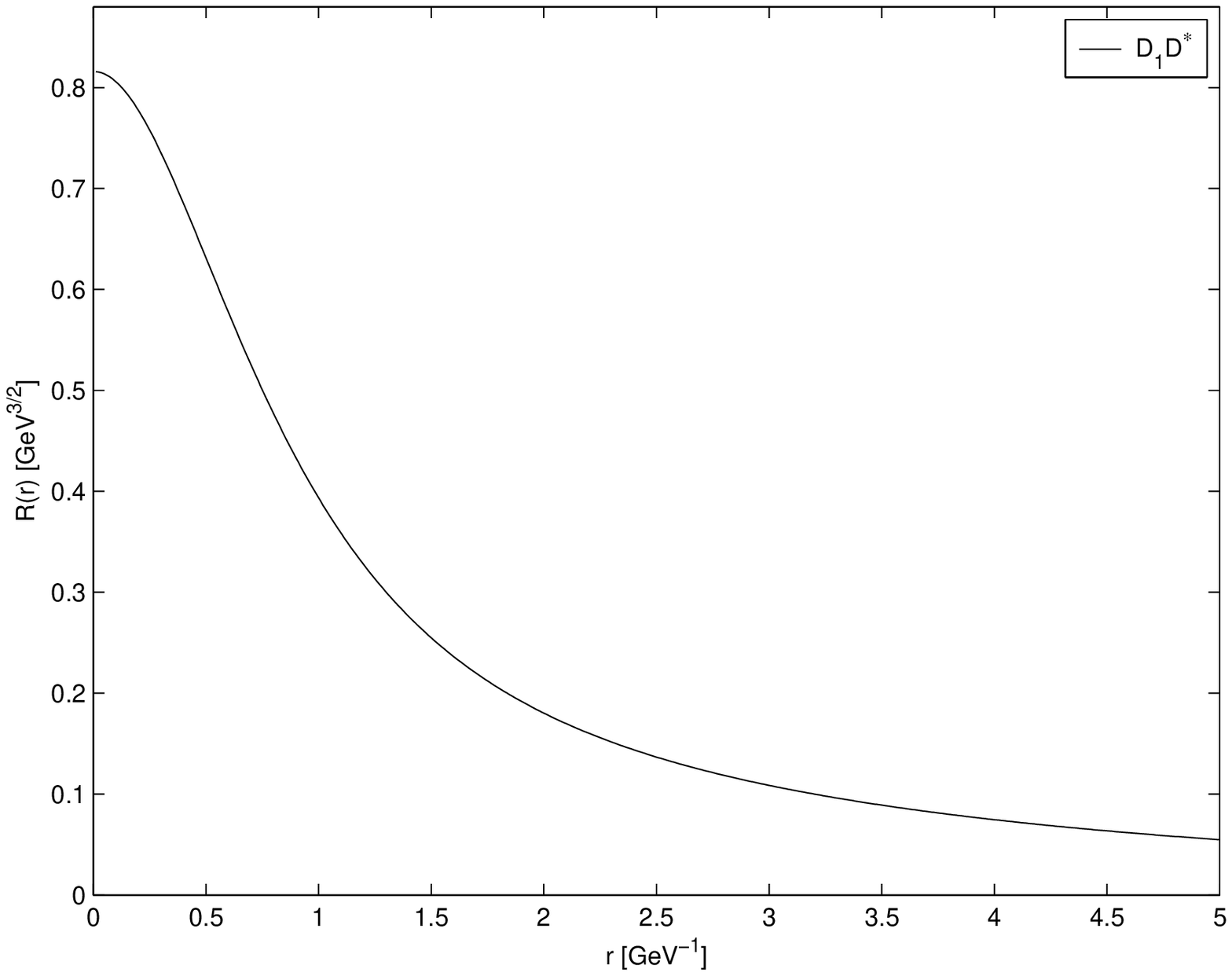}}\\
(a)&(b)
\end{tabular}
\caption{The radial wave function $R(r)$ for the molecular state
with $J=0$ of the $D_1'D^*$ and $D_1D^*$ system respectively. The
corresponding parameters are [$g\cdot g'$ = 0.1, $h$ = 0.84,
$\Lambda$=1.5 GeV] and [$g\cdot g"$ = 0.2, $h'$ = 0.55 GeV$^{-1}$,
$\Lambda$=2.9 GeV] respectively.\label{wave-R}}
\end{center}
\end{figure}

\begin{figure}[htb]
\begin{center}
\begin{tabular}{ccc}
\scalebox{0.32}{\includegraphics{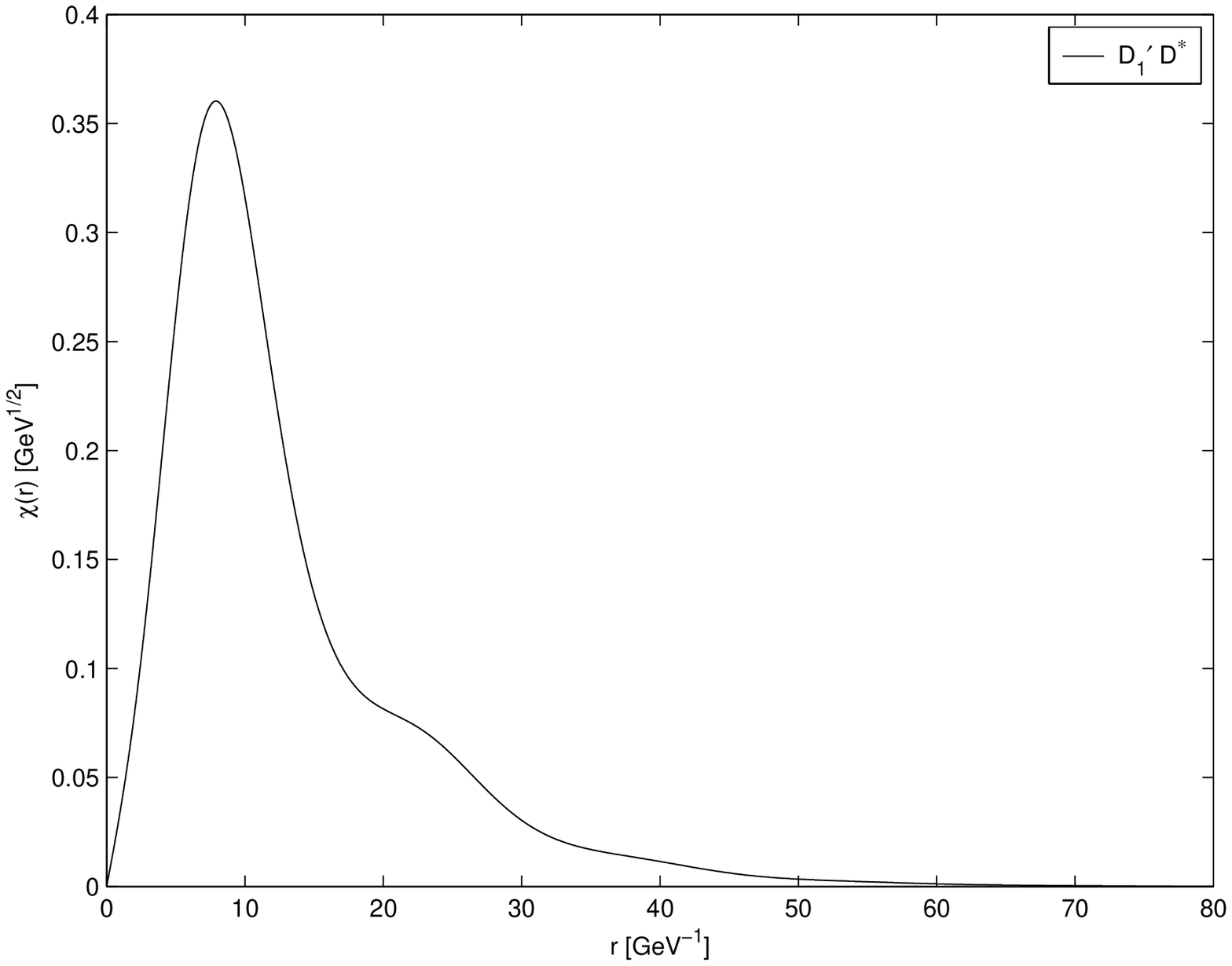}}&
\scalebox{0.32}{\includegraphics{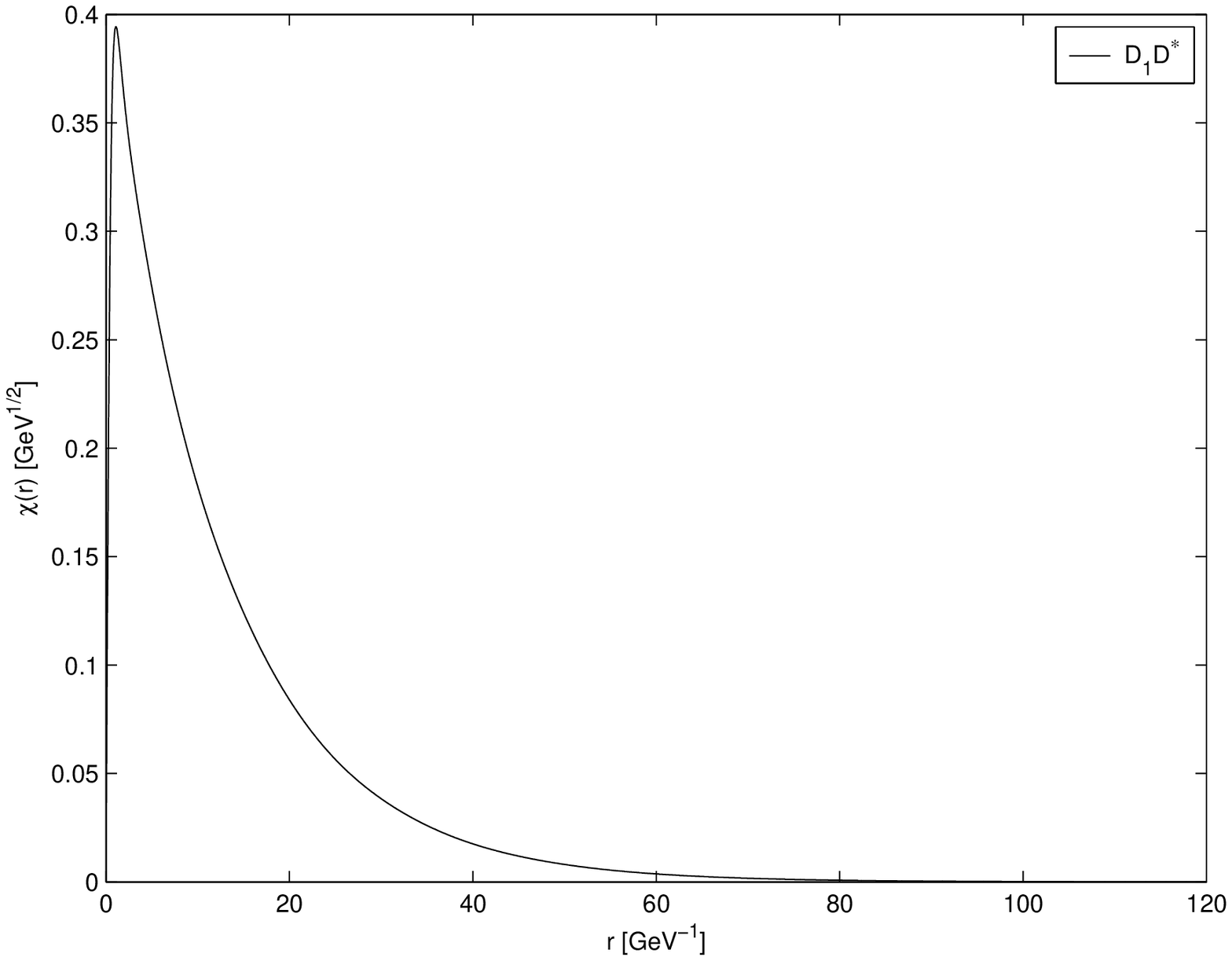}}&
\scalebox{0.32}{\includegraphics{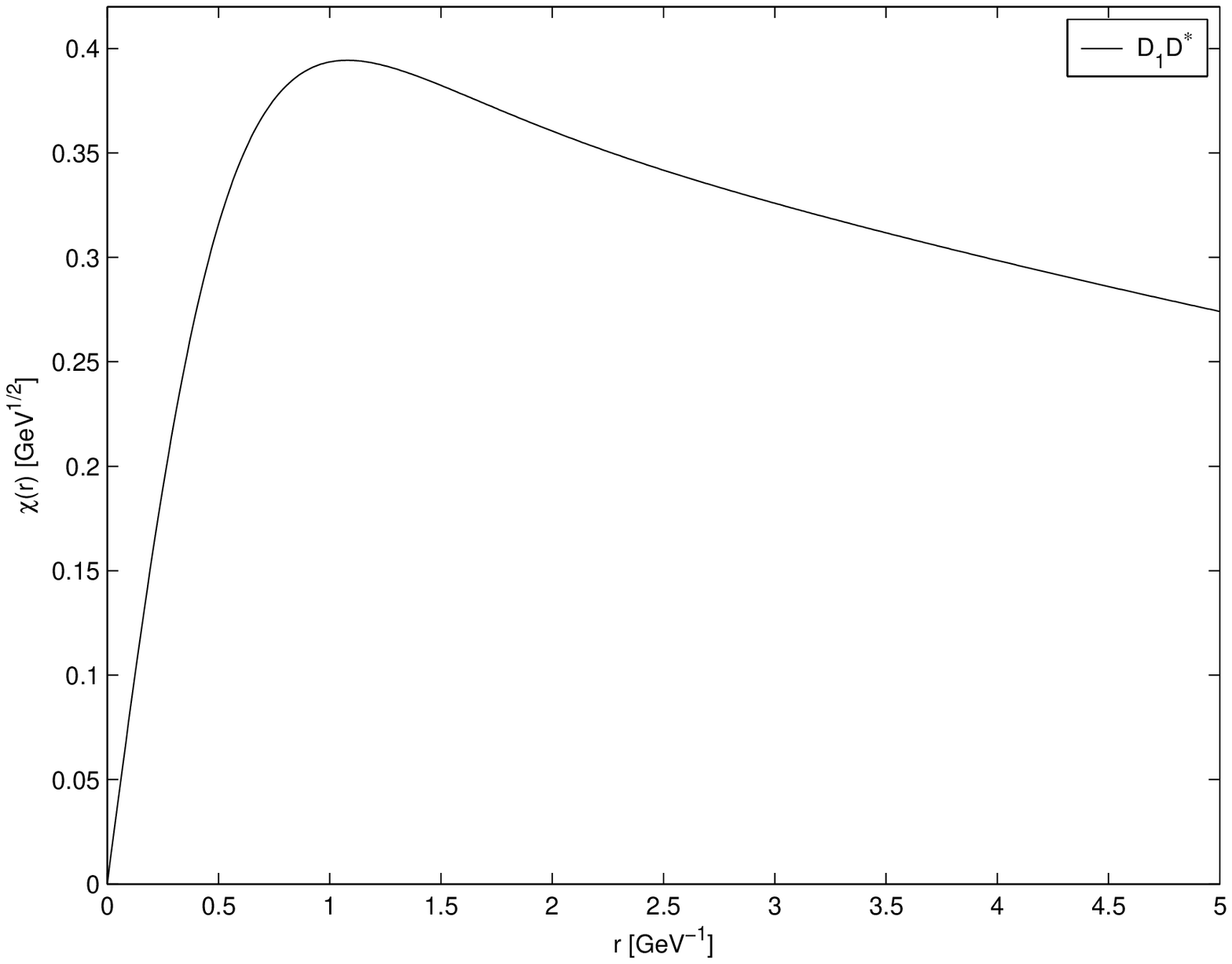}}\\
(a)&(b)&(c)
\end{tabular}
\caption{The function $\chi(r)=r R(r)$ for the J=0 molecular state
of the $D_1'D^*$ and $D_1D^*$ system. (c) shows the short range
behavior for the $D_1D^*$ case. The corresponding parameters for
the two systems are [$g\cdot g'$ = 0.1, $h$ = 0.84, $\Lambda$=1.5
GeV] and [$g\cdot g''$ = 0.2, $h'$ = 0.55 GeV$^{-1}$,
$\Lambda$=2.9 GeV] respectively. \label{wave-chi}}
\end{center}
\end{figure}

In Figs. \ref{wave-R} (a) and \ref{wave-chi} (a) we show the
radial wave function $R(r)$ and function $\chi(r)=r R(r)$ for the
$D_1'-D^*$ system with $J=0$. We list the numerical results with
different typical values of coupling constants in Table
\ref{bind-D1prime} for the $D_{1}'-D^*$ system. Here $r_{rms}$ is
the root-mean-square radius and $r_{max}$ denotes the radius
corresponding to the maximum of the wave function $\chi(r)$ of
$D_1'-D^*$ system. $E(\Lambda)$ denotes the binding energy with
the corresponding cutoff. For example, the notation -6.0(1.5)
denotes the binding energy is 6.0 MeV at the cutoff $\Lambda=1.5$
GeV.

From the numerical results listed in Table \ref{bind-D1prime}, one
concludes that the existence of S-wave $D_{1}'-D^*$ bound state
with $J^P=0^-,1^-,2^-$ is possible. With appropriate parameters,
one can get a molecular state consistent with $Z^+(4430)$.
Throughout our study, we have ignored the width of heavy mesons.
However, the broad width of $D_1'$ around $\Gamma\sim$384 MeV
\cite{PDG} may be a obstacle for the formation of the molecular
state, which deserves further study.

By comparing the results with different sets of parameters, one
finds that the sigma exchange interaction induces very small
effects on the binding energy. Of the parameters $g\cdot g'$ and
$h$, the binding energy is sensitive to $h$, which indicates that
the crossing diagram from one pion exchange plays an important
role in binding $D_1'D^*$. Numerically, large $h$, small $g\cdot
g'$ for $J=0,1$ and big $g\cdot g'$ for $J=2$ are helpful to form
a bound state. These observations are consistent with the
conclusions by analyzing the dependence of the potentials on
different coupling constants in the previous section.

In Table \ref{bind-D1prime}, we only give results corresponding to
two cutoffs $\Lambda=0.7$ GeV and $\Lambda=1.5$ GeV. By comparing
the binding energies with different $\Lambda$ when $h=0.84$, one
notes that $E$ becomes larger with a smaller $\Lambda$. We come
back to this point later when discussing the cutoff dependence of
the binding energy $E$.

\begin{center}
\begin{table}
\begin{tabular}{c|c|c|c||ccc|ccc|ccc}\hline
\multicolumn{13}{c}{$D_1'-D^*$ system}\\\hline\hline
\multicolumn{4}{c}{}&\multicolumn{3}{c}{$J=0$}&\multicolumn{3}{c}{$J=1$}&\multicolumn{3}{c}{$J=2$}\\\hline
$g\cdot g'$&$h$&$g_{\sigma}\cdot
g_{\sigma}'$&$h_{\sigma}$&$E(\Lambda)$&$r_{rms}$&$r_{max}$&$E(\Lambda)$&$r_{rms}$&$r_{max}$
&$E(\Lambda)$&$r_{rms}$&$r_{max}$\\\hline

0.1&0.56&-&-&-2.8(0.7)&2.5&1.5&-2.9(0.7)&2.4&1.5&-3.1(0.7)&2.4&1.4
\\\hline
&&&0.4&-2.8(0.7)&2.5&1.5&-2.9(0.7)&2.4&1.5&-3.1(0.7)&2.4&1.4\\
&&\raisebox{1.5ex}{0.58}&0.8&-2.8(0.7)&2.4&1.5&-2.9(0.7)&2.4&1.5&-3.1(0.7)&2.4&1.4\\\cline{3-13}
\raisebox{2ex}{0.1}&\raisebox{2ex}{0.56}&&0.4&-2.8(0.7)&2.5&1.5&-2.9(0.7)&2.4&1.5&-3.1(0.7)&2.4&1.5\\
&&\raisebox{1.5ex}{-0.58}&0.8&-2.8(0.7)&2.5&1.5&-2.9(0.7)&2.4&1.5&-3.1(0.7)&2.4&1.4\\\hline\hline

0.1&0.84&-&-&-23.9(0.7)/-5.5(1.5)&1.5/2.1&1.3/1.5&-24.1(0.7)/-5.6(1.5)&1.5/2.1&1.3/1.5&-24.5(0.7)/-5.8(1.5)&1.5/2.1&1.2/1.5
\\\hline
&&&0.4&-23.9(0.7)/-5.4(1.5)&1.5/2.1&1.3/1.5&-24.1(0.7)/-5.5(1.5)&1.5/2.1&1.3/1.5&-24.5(0.7)/-5.6(1.5)&1.5/2.1&1.2/1.5\\
&&\raisebox{1.5ex}{0.58}&0.8&-24.1(0.7)/-5.5(1.5)&1.5/2.0&1.3/1.5&-24.1(0.7)/-5.5(1.5)&1.5/2.1&1.3/1.5&-24.5(0.7)/-5.7(1.5)&1.5/2.1&1.2/1.5\\\cline{3-13}
\raisebox{2ex}{0.1}&\raisebox{2ex}{0.84}&&0.4&-23.9(0.7)/-5.6(1.5)&1.5/2.1&1.3/1.5&-24.1(0.7)/-5.7(1.5)&1.5/2.1&1.3/1.5&-24.5(0.7)/-5.9(1.5)&1.5/2.0&1.2/1.5\\
&&\raisebox{1.5ex}{-0.58}&0.8&-23.9(0.7)/-5.9(1.5)&1.5/2.0&1.3/1.5&-24.1(0.7)/-5.8(1.5)&1.5/2.0&1.3/1.5&-24.5(0.7)/-5.9(1.5)&1.5/2.0&1.2/1.5\\\hline\hline

0.5&0.56&-&-&-2.0(0.7)&2.8&1.5&-2.4(0.7)&2.6&1.5&-3.6(0.7)&2.2&1.4\\\hline
&&&0.4&-2.0(0.7)&2.8&1.5&-2.5(0.7)&2.6&1.5&-3.6(0.7)&2.2&1.4\\
&&\raisebox{1.5ex}{0.58}&0.8&-2.0(0.7)&2.8&1.5&-2.5(0.7)&2.6&1.5&-3.5(0.7)&2.2&1.4\\\cline{3-13}
\raisebox{2ex}{0.5}&\raisebox{2ex}{0.56}&&0.4&-2.0(0.7)&2.8&1.5&-2.5(0.7)&2.6&1.5&-3.6(0.7)&2.2&1.4\\
&&\raisebox{1.5ex}{-0.58}&0.8&-2.0(0.7)&2.8&1.5&-2.5(0.7)&2.6&1.5&-3.6(0.7)&2.2&1.4\\\hline\hline

0.5&0.84&-&-&-22.3(0.7)/-5.0(1.5)&1.5/2.2&1.3/1.6&-23.3(0.7)/-5.3(1.5)&1.5/2.1&1.3/1.5&-25.3(0.7)/-6.2(1.5)&1.4/2.0&1.2/1.5\\\hline
&&&0.4&-22.3(0.7)/-4.8(1.5)&1.5/2.2&1.3/1.6&-23.3(0.7)/-5.2(1.5)&1.5/2.1&1.3/1.5&-25.3(0.7)/-6.0(1.5)&1.4/2.0&1.2/1.5\\
&&\raisebox{1.5ex}{0.58}&0.8&-22.3(0.7)/-4.7(1.5)&1.5/2.2&1.3/1.6&-23.3(0.7)/-5.1(1.5)&1.5/2.1&1.3/1.5&-25.3(0.7)/-6.0(1.5)&1.4/2.0&1.2/1.5\\\cline{3-13}
\raisebox{2ex}{0.5}&\raisebox{2ex}{0.84}&&0.4&-22.3(0.7)/-5.1(1.5)&1.5/2.2&1.3/1.6&-23.3(0.7)/-5.4(1.5)&1.5/2.1&1.3/1.5&-25.3(0.7)/-6.3(1.5)&1.4/2.0&1.2/1.5\\
&&\raisebox{1.5ex}{-0.58}&0.8&-22.4(0.7)/-5.0(1.5)&1.5/2.1&1.3/1.5&-23.3(0.7)/-5.4(1.5)&1.5/2.1&1.3/1.5&-25.3(0.7)/-6.3(1.5)&1.4/2.0&1.2/1.5\\\hline\hline

-0.1&0.56&-&-&-3.2(0.7)&2.3&1.4&-3.1(0.7)&2.3&1.4&-2.9(0.7)&2.4&1.4
\\\hline
&&&0.4&-3.2(0.7)&2.3&1.4&-3.1(0.7)&2.4&1.4&-2.9(0.7)&2.4&1.5\\
&&\raisebox{1.5ex}{0.58}&0.8&-3.2(0.7)&2.3&1.4&-3.1(0.7)&2.4&1.4&-2.9(0.7)&2.4&1.5\\\cline{3-13}
\raisebox{2ex}{-0.1}&\raisebox{2ex}{0.56}&&0.4&-3.2(0.7)&2.3&1.4&-3.1(0.7)&2.4&1.4&-2.9(0.7)&2.4&1.5\\
&&\raisebox{1.5ex}{-0.58}&0.8&-3.2(0.7)&2.3&1.4&-3.1(0.7)&2.4&1.4&-2.9(0.7)&2.4&1.5\\\hline\hline

-0.1&0.84&-&-&-24.7(0.7)/-5.9(1.5)&1.4/2.0&1.2/1.5&-24.5(0.7)/-5.8(1.5)&1.5/2.0&1.2/1.5&-24.1(0.7)/-5.6(1.5)&1.5/2.1&1.3/1.5
\\\hline
&&&0.4&-24.7(0.7)/-5.7(1.5)&1.4/2.0&1.2/1.5&-24.5(0.7)/-5.6(1.5)&1.5/2.1&1.2/1.5&-24.1(0.7)/-5.5(1.5)&1.5/2.1&1.3/1.5\\
&&\raisebox{1.5ex}{0.58}&0.8&-24.7(0.7)/-6.5(1.5)&1.4/1.9&1.2/1.5&-24.5(0.7)/-6.0(1.5)&1.5/2.0&1.2/1.5&-24.1(0.7)/-5.8(1.5)&1.5/2.1&1.3/1.5\\\cline{3-13}
\raisebox{2ex}{-0.1}&\raisebox{2ex}{0.84}&&0.4&-24.7(0.7)/-6.0(1.5)&1.4/2.0&1.2/1.5&-24.5(0.7)/-5.9(1.5)&1.5/2.0&1.2/1.5&-24.1(0.7)/-5.7(1.5)&1.5/2.1&1.3/1.5\\
&&\raisebox{1.5ex}{-0.58}&0.8&-24.7(0.7)/-6.5(1.5)&1.4/1.9&1.2/1.5&-24.5(0.7)/-6.0(1.5)&1.5/2.0&1.2/1.5&-24.1(0.7)/-5.8(1.5)&1.5/2.1&1.3/1.5\\\hline\hline

-0.5&0.56&-&-&-4.2(0.7)&2.1&1.4&-3.6(0.7)&2.2&1.4&-2.4(0.7)&2.6&1.5\\\hline
&&&0.4&-4.2(0.7)&2.1&1.4&-3.6(0.7)&2.2&1.4&-2.4(0.7)&2.6&1.5\\
&&\raisebox{1.0ex}{0.58}&0.8&-4.2(0.7)/-1.1(1.5)&2.1/3.3&1.4/1.4&-3.6(0.7)&2.2&1.4&-2.4(0.7)&2.6&1.5\\\cline{3-13}
\raisebox{2ex}{-0.5}&\raisebox{2ex}{0.56}&&0.4&-4.2(0.7)/-0.1(1.5)&2.1/10.6&1.4/1.7&-3.6(0.7)&2.2&1.4&-2.4(0.7)&2.6&1.5\\
&&\raisebox{1.0ex}{-0.58}&0.8&-4.2(0.7)/-1.6(1.5)&2.1/2.8&1.4/1.4&-3.6(0.7)&2.2&1.4&-2.4(0.7)&2.6&1.5\\\hline\hline

-0.5&0.84&-&-&-26.4(0.7)/-6.8(1.5)&1.4/2.0&1.2/1.5&-25.3(0.7)/-6.2(1.5)&1.4/2.0&1.2/1.5&-23.3(0.7)/-7.6(1.5)&1.5/2.1&1.3/1.5\\\hline
&&&0.4&-26.4(0.7)/-6.8(1.5)&1.4/1.9&1.2/1.5&-25.3(0.7)/-6.0(1.5)&1.4/2.0&1.2/1.5&-23.3(0.7)/-5.2(1.5)&1.5/2.1&1.3/1.5\\
&&\raisebox{1.5ex}{0.58}&0.8&-26.4(0.7)/-8.0(1.5)&1.4/1.8&1.2/1.4&-25.3(0.7)/-6.2(1.5)&1.4/2.0&1.2/1.5&-23.3(0.7)/-5.3(1.5)&1.5/2.1&1.3/1.5\\\cline{3-13}
\raisebox{2ex}{-0.5}&\raisebox{2ex}{0.84}&&0.4&-26.4(0.7)/-7.2(1.5)&1.4/1.9&1.2/1.5&-25.3(0.7)/-6.4(1.5)&1.4/2.0&1.2/1.5&-23.3(0.7)/-5.5(1.5)&1.5/2.1&1.3/1.5\\
&&\raisebox{1.5ex}{-0.58}&0.8&-26.4(0.7)/-8.6(1.5)&1.4/1.8&1.2/1.4&-25.3(0.7)/-6.6(1.5)&1.4/1.9&1.2/1.5&-23.3(0.7)/-5.6(1.5)&1.5/2.1&1.3/1.5\\\hline\hline

\end{tabular}
\caption{The numerical results of $D_1'-D^*$ system with $J^P=0^-,
1^-, 2^-$. Here the bing energy $E$, cutoff $\Lambda$, the
root-mean-square radius $r_{rms}$ and $r_{max}$ of the $D_1'-D^*$
system are in unit of MeV, GeV, fm and fm respectively.
$E(\Lambda)$ denotes the binding energy with the corresponding
cutoff. \label{bind-D1prime}}
\end{table}
\end{center}

\subsection{S-wave $D_{1}-D^*$ system}

In Table \ref{bind-D1}, we present the numerical results for the
case of S-wave $D_1-D^*$ system. Unfortunately one fails to find
solutions with negative binding energy for $J=1,2$ using the
parameters in that table, which indicates that there probably does
not exist the S-wave $D_1-D^*$ molecule with $J^P=1^-,2^-$ through
the pion and sigma exchange interactions. However the S-wave
$D_1-D^*$ molecular state with $J^P=0^-$ probably exists. Thus we
only present the results for the $D_1-D^*$ system with $J=0$.

When taking cutoff $\Lambda=2.9$ GeV, one gets $E=-1.3$ MeV with
parameters $[g\cdot g'', h', g_{\sigma}\cdot g_{\sigma}'',
h_{\sigma}'$]=[0.2,0.55 GeV$^{-1}$,0.58,0.2]. Varying parameters
$g\cdot g''$, $g_{\sigma}\cdot g_{\sigma}''$, $h_{\sigma}'$ in the
reasonable range results in large change for the binding energy
with fixed $\Lambda=2.9$ GeV, which indicates that the solutions
are sensitive to coupling constants. Meanwhile one finds that one
sigma exchange interaction induces significant effects. There
exist solutions with the binding energies around $-10\sim 0$ MeV
with appropriate coupling constants and cutoff, which are also
shown in Table \ref{bind-D1}. We present the radial wave function
$R(r)$ and function $\chi(r)=r R(r)$ for the $D_1-D^*$ system with
$J=0$ in Fig. \ref{wave-R} (b) and Fig. \ref{wave-chi} (b), (c)
respectively.

One finds that big $g\cdot g''$, small $g_\sigma\cdot g_\sigma''$
and big $h_\sigma'$ are beneficial to large binding energy. The
results indicate that the larger the cutoff $\Lambda$ is, the
deeper the binding.

\begin{center}
\begin{table}[htb]
\begin{tabular}{c|c|c|c||ccc}\hline
\multicolumn{7}{c}{$D_1-D^*$ system}\\\hline\hline
\multicolumn{4}{c}{}&\multicolumn{3}{c}{$J=0$}\\\hline $g\cdot
g''$&$h'$&$g_{\sigma}\cdot
g_{\sigma}''$&$h_{\sigma}'$&E($\Lambda$)&$r_{rms}$(fm)&$r_{max}$(fm)\\\hline
&&0.0&0.0&-2.8(2.9)&1.8&0.2\\
&&&0.2&-1.3(2.9)&2.6&0.2\\
&&\raisebox{0.5ex}{0.58}&1.0&-133.2(2.9)&0.3&0.2\\
\raisebox{2ex}{0.2}&\raisebox{2ex}{0.55}&&0.2&-9.8(2.9)&1.0&0.2\\
&&\raisebox{0.5ex}{-0.58}&1.0&-155.5(2.9)&0.3&0.2\\\hline
&&0.0&0.0&-196.4(2.9)&0.3&0.2\\
&&&0.2&-193.3(2.9)&0.3&0.2\\
&&\raisebox{0.5ex}{0.58}&1.0&-4.6(2.0)/-427.0(2.9)&1.5/0.2&0.4/0.1\\
\raisebox{2ex}{0.6}&\raisebox{2ex}{0.55}&&0.2&-217.0(2.9)&0.3&0.2\\
&&\raisebox{0.5ex}{-0.58}&1.0&-9.1(2.0)/-453.6(2.9)&1.1/0.2&0.4/0.1\\\hline
&&0.0&0.0&$\times$&$\times$&$\times$\\
&&&0.2&$\times$&$\times$&$\times$\\
&&\raisebox{0.5ex}{0.58}&1.0&$\times$&$\times$&$\times$\\
\raisebox{2ex}{-0.2}&\raisebox{2ex}{0.55}&&0.2&$\times$&$\times$&$\times$\\
&&\raisebox{0.5ex}{-0.58}&1.0&$\times$&$\times$&$\times$\\\hline
&&0.0&0.0&$\times$&$\times$&$\times$\\
&&&0.2&$\times$&$\times$&$\times$\\
&&\raisebox{0.5ex}{0.58}&1.0&$\times$&$\times$&$\times$\\
\raisebox{2ex}{-0.6}&\raisebox{2ex}{0.55}&&0.2&$\times$&$\times$&$\times$\\
&&\raisebox{0.5ex}{-0.58}&1.0&$\times$&$\times$&$\times$\\\hline
\end{tabular}
\caption{Numerical results for the $D_1-D^*$ system with several
sets of parameters if we use $\Lambda=2.0$ GeV and $\Lambda=2.9$
GeV. The unit is MeV. Negative bound energies exist only for
$J=0$. The cross $\times$ means no bound state exists. If we use a
smaller cutoff $\Lambda=0.7$ GeV, there are no bound state
solutions with the parameters in this table.\label{bind-D1}}
\end{table}
\end{center}

\section{Bottom analog }\label{sec6}

The calculation can be easily extended to study the bottom analog
of $Z^+(4430)$. For such a system, its flavor wavefunction is
\begin{eqnarray}\label{Bwave-1}
|Z_B^{+}\rangle=\frac{1}{\sqrt{2}}\Big [|B_{1}^{'+}
\bar{B}^{*0}\rangle+|B^{*+}\bar{B}_{1}^{'0}\rangle\Big]
\end{eqnarray} or
\begin{eqnarray}\label{Bwave-2}
|Z_B^{+}\rangle=\frac{1}{\sqrt{2}}\Big [|B_{1}^{+}
\bar{B}^{*0}\rangle+|B^{*+}\bar{B}_{1}^{0}\rangle\Big].
\end{eqnarray}

\begin{center}
\begin{table}[htb]
\begin{tabular}{c|c|ccc||ccc||ccc}\hline
\multicolumn{11}{c}{$B_1'-B^*$ system}\\\hline\hline
\multicolumn{2}{c}{}&\multicolumn{3}{c}{$J=0$}&\multicolumn{3}{c}{$J=1$}&\multicolumn{3}{c}{$J=2$}\\\hline
$g\cdot
g'$&$h$&E(MeV)&$r_{rms}$(fm)&$r_{max}$(fm)&E(MeV)&$r_{rms}$(fm)&$r_{max}$(fm)&E(MeV)&$r_{rms}$(fm)&$r_{max}$(fm)\\\hline
0.1&0.56&-2.6&2.0&1.6&-2.6&2.0&1.6&-2.8&2.0&1.6\\
0.1&0.84&-14.6&1.6&1.5&-14.7&1.5&1.5&-14.7&1.6&1.5\\
0.5&0.56&-2.3&2.1&1.6&-2.5&2.1&1.6&-3.2&1.9&1.5\\
0.5&0.84&-14.5&1.7&1.5&-14.6&1.7&1.5&-14.9&1.6&1.5\\
-0.1&0.56&-2.9&2.0&1.6&-2.8&2.0&1.6&-2.6&2.0&1.6\\
-0.1&0.84&-14.8&1.6&1.5&-14.7&1.6&1.5&-14.7&1.6&1.5\\
-0.5&0.56&-4.4&1.7&1.4&-3.2&1.9&1.5&-2.5&2.1&1.6\\
-0.5&0.84&-15.3&1.6&1.5&-14.9&1.6&1.5&-14.6&1.7&1.5\\\hline
\end{tabular}
\caption{Numerical results for the $B_1'-B^*$ system with several
sets of parameters and $\Lambda=1.5$ GeV. \label{Banalog-1}}
\end{table}
\end{center}

\begin{center}
\begin{table}[htb]
\begin{tabular}{c|c|c|c||ccc}\hline
\multicolumn{7}{c}{$B_1-B^*$ system}\\\hline\hline
\multicolumn{4}{c}{}&\multicolumn{3}{c}{$J=0$}\\\hline $g\cdot
g''$&$h'$&$g_{\sigma}\cdot
g_{\sigma}''$&$h_{\sigma}'$&E($\Lambda$)&$r_{rms}$(fm)&$r_{max}$(fm)\\\hline
&&0.0&0.0&-1.0(1.9)&1.9&0.4\\
&&&0.2&-0.2(1.9)&4.3&0.4\\
&&\raisebox{0.5ex}{0.58}&1.0&-26.9(1.9)&0.5&0.3\\
\raisebox{2ex}{0.2}&\raisebox{2ex}{0.55}&&0.2&-3.6(1.9)&1.1&0.3\\
&&\raisebox{0.5ex}{-0.58}&1.0&-35.3(1.9)&0.5&0.2\\\hline
&&0.0&0.0&-0.2(1.2)/-60.7(1.9)&4.4/0.4&0.7/0.2\\
&&&0.2&-0.1(1.2)/-58.1(1.9)&7.5/0.4&0.7/0.2\\
&&\raisebox{0.5ex}{0.58}&1.0&-1.3(1.2)/-111.1(1.9)&1.8/0.3&0.6/0.2\\
\raisebox{2ex}{0.6}&\raisebox{2ex}{0.55}&&0.2&-0.4(1.2)/-67.4(1.9)&3.0/0.4&0.7/0.2\\
&&\raisebox{0.5ex}{-0.58}&1.0&-2.1(1.2)/-121.4(1.9)&1.5/0.3&0.6/0.2\\\hline
&&0.0&0.0&$\times$&$\times$&$\times$\\
&&&0.2&$\times$&$\times$&$\times$\\
&&\raisebox{0.5ex}{0.58}&1.0&$\times$&$\times$&$\times$\\
\raisebox{2ex}{-0.2}&\raisebox{2ex}{0.55}&&0.2&$\times$&$\times$&$\times$\\
&&\raisebox{0.5ex}{-0.58}&1.0&$\times$&$\times$&$\times$\\\hline
&&0.0&0.0&$\times$&$\times$&$\times$\\
&&&0.2&$\times$&$\times$&$\times$\\
&&\raisebox{0.5ex}{0.58}&1.0&$\times$&$\times$&$\times$\\
\raisebox{2ex}{-0.6}&\raisebox{2ex}{0.55}&&0.2&$\times$&$\times$&$\times$\\
&&\raisebox{0.5ex}{-0.58}&1.0&$\times$&$\times$&$\times$\\\hline
\end{tabular}
\caption{Numerical results for the $B_1-B^*$ system with several
sets of parameters if we use $\Lambda=1.2$ GeV and $\Lambda=1.9$
GeV. Here $E(\Lambda)$ denotes the binding energy with
corresponding cutoff. The unit is MeV. Negative bound energies
exist only for $J=0$. \label{Banalog-2}}
\end{table}
\end{center}

The masses of the bottom mesons are $m_{B^\ast}=5325.0$ MeV,
$m_{B_1}=5725.3$ MeV and $m_{B_1'}=5732$ MeV \cite{PDG,PDG-1}.
Since $m_{B_{1}'}-m_{B^\ast}$ or $m_{B_{1}}-m_{B^\ast}$ is greater
than the pion mass and less than $\sigma$ mass, the forms of the
potentials are exactly those given in Table \ref{table-1} and
\ref{table-2}. The difference between the charmed and bottom
systems lies only in the meson masses. The relatively small
kinematic term makes it easier to form molecular states in the
bottom system.

From the results for the $D_1'D^\ast$ system (Table
\ref{bind-D1prime}), one sees that the $\sigma$ exchange gives
negligible contributions to the binding energy of the system. The
same conclusion holds for the $B_1'B^\ast$ system. Thus we ignore
effects from $\sigma$ exchange in the numerical evaluation. We
choose eight sets of parameters to calculate: [$g\cdot g'$,
$h$]=[$\pm0.1$, 0.56], [$\pm0.5$, 0.56], [$\pm0.1$, 0.84] and
[$\pm0.5$, 0.84]. The results are given in Table \ref{Banalog-1}
where we use the cutoff $\Lambda=1.5$ GeV. With this cutoff, a
bound state always exists with reasonable couplings.

For the system of $B_1B^\ast$, we use the same sets of parameters
in Table \ref{bind-D1} since the contribution from the $\sigma$
exchange may be large. We list the results in Table
\ref{Banalog-2} where two values of $\Lambda$ 1.2 GeV and 1.9 GeV,
are used. This case is similar to the case of the $D_1D^\ast$
case. The bound states with $J=0$ may exist with the positive
$g\cdot g''$.

\section{The cutoff dependence}\label{sec7}

Up to now, we have solved the radial Schroding equation with only
a few values of the cutoff $\Lambda$. But the binding energies
will change with the variation of this parameter. We briefly study
the cutoff dependence of the binding energy.

Since a bound state is more easily formed in the bottom system, we
first study the system $B_1'B^\ast$ with $g\cdot g'=-0.5$ and
$h=0.84$ for $J=0$. By scanning results starting from
$\Lambda=0.7$ GeV, we find the binding energy has a smallest value
around $\Lambda=2.3$ GeV. One can always find negative eigenvalues
with $g\cdot g'=-0.5$ and $h=0.84$ for this case. The dependence
of the binding energies on the cutoff is not monotonic. We also
study results with other couplings and $J$. This conclusion is
always correct for the $B_1'B^\ast$ system. Some minimum binding
energies with corresponding $\Lambda$ are presented in Table
\ref{BdepC}.

\begin{table}[htb]
\centering
\begin{tabular}{c||cccc|ccc}\hline
Systems&$g\cdot g'$&$h$&$g_\sigma\cdot
g_\sigma'$&$h_\sigma$&$J=0$&$J=1$&$J=2$\\\hline\hline
$B_1'{B}^\ast$&-0.5&0.84&0&0&-14.4(2.3)&-14.0(3.5)&-14.4(2.1)\\
&0.5&0.56&0&0&-2.3(1.4)&-2.4(1.8)&-2.8(2.4)\\\hline
$D_1'{D}^\ast$&-0.5&0.84&-0.58&0.8&-8.6(1.5)&-5.9(2.1)&-5.5(1.7)\\\hline
\end{tabular}
\caption{Minimum binding energies $E$ with the corresponding
cutoff $\Lambda$ for some sets of coupling constants. The unit for
$E(\Lambda)$ is MeV(GeV).}\label{BdepC}
\end{table}

It is interesting to study whether similar behavior exists in the
other systems. For the $D_1'{D}^\ast$ system, the minimum binding
energy exists only for some coupling constants. One example is
presented in Table \ref{BdepC}. For the solutions with $h=0.56$,
the binding energy decreases if we increase $\Lambda$ until the
$D_1'{D}^\ast$ pair is no longer bound. Thus bound states exist
only in a small range of $\Lambda$.

For the $D_1{D}^\ast$ and $B_1{B}^\ast$ systems, the solutions for
bound states can be found only when $\Lambda$ is big enough. The
binding energy becomes large with the increase of the cutoff.

In short summary, we have studied the cutoff dependence for the
binding energies with some sets of coupling constants. Three types
of behavior are found: (1) bound state solutions always exist
whatever the $\Lambda$ is. In this case, the binding energy
reaches the minimum value at a special $\Lambda$; (2) bound state
solutions exist with small $\Lambda$ only. The binding energy
becomes large when the cutoff becomes small; (3) bound state
solutions exist when $\Lambda$ is big enough. In this case, the
binding energy increases with the cutoff. Which case is realized
for a system depends on the properties of the components and the
values of the coupling constants. One should also keep in mind the
cutoff is a typical hadronic scale related to the system.

\section{Conclusions}\label{sec8}

QCD allows the possible existence of the glueball, hybrids,
multiquarks and molecular states etc. However, none of them has
been established firmly. Recently Belle collaboration announced
the observation of charged enhancement $Z^+(4430)$, whose unique
properties make $Z^+(4430)$ hardly understood as a conventional
meson. A natural explanation of $Z^+(4430)$ is the $D_1^{'}D^*$ or
$D_1D^*$ molecule state.

In this work, we re-examine the dynamics of $Z^+(4430)$ and
improve the analysis in Ref. \cite{xiangliu} in the following
aspects: (1) we include the sigma meson exchange contribution
besides the one pion exchange potential; (2) we introduce the form
factor to take into account the structure effect of the
interaction vertex; (3) we solve the schr\"{o}dinger equation of
the S-wave $D_1^{'}D^*$ or $D_1D^*$ system with the help of Matlab
package MATSLISE.

We find that the one pion exchange potential from the crossed
diagram plays a dominant role for the S-wave $D_1'D^*$ or $D_1
D^*$ system. Our numerical results indicate that with the coupling
constants determined in Section \ref{sec2} there exists the S-wave
$D_1'D^*$ molecular state with $J^{P}=0^-,1^-,2^-$. The sigma
meson exchange contribution to the binding energy is small.
However one should carefully study whether the broad width of
$D_1'$ disfavors the formation of a molecular state in the future.

For the S-wave $D_1-D^*$ system, only a molecular state with
$J^P=0^-$ may exist with appropriate parameters. The contribution
from the sigma meson exchange is significant in this case.
Replacing the charmed meson masses with bottom meson masses, we
have also studied the bottom analogy $Z_B$ of the $Z(4430)$
system. As expected, the absolute value of the binding energy of
$Z_B$ is larger than that of $Z(4430)$. Such a state may be
searched for at Tevatron and LHC. Clearly future experimental
measurement of the quantum numbers of $Z(4430)$ will help uncover
its underlying structure.

\section*{Acknowledgments}

We thank Professor K.T. Chao, Professor E. Braaten and Professor
Z.Y. Zhang for useful discussions. This project was supported by
National Natural Science Foundation of China under Grants
10625521, 10675008, 10705001, 10775146, 10721063 and China
Postdoctoral Science foundation (20060400376, 20070420526). X.L.
was partly supported by the \emph{Funda\c{c}\~{a}o para a
Ci\^{e}ncia e a Tecnologia of the Minist\'{e}rio da Ci\^{e}ncia,
Tecnologia e Ensino Superior} of Portugal (SFRH/BPD/34819/2007).

\end{document}